# Metric-driven search for structurally stable inorganic compounds

R. Villarreal,[a,b*] P. Singh,[a,*,+] and, R. Arroyave[a,b,c]

[a]Department of Materials Science & Engineering, Texas A&M University, College Station, TX 77843, USA
[b] Department of Mechanical Engineering, Texas A&M University, College Station, TX 77843, USA
[c] Department of Industrial and Systems Engineering, Texas A&M University, College Station, TX 77843, USA

* These authors equally contributed to the execution of the present work
+ Current address: Ames Laboratory, U.S. Department of Energy, Iowa State University, Ames, Iowa 50011 USA

## Abstract

We report a facile `metric' for the identification of structurally and dynamically (positive definite phonon structure) stable inorganic compounds. The metric considers charge-imbalance within the local substructures in crystalline compounds calculated using first-principles density-functional theory. To exemplify, we chose carbon-based nitrides as it provides a large pool of structurally stable and unstable phases. We showcase how local structural information related to Wyckoff symmetry uniquely identifies four new carbon-nitride phases. The metric predicts three new structurally stable phases of 4 (C):3 (N) stoichiometry, i.e., $C_4N_3$, which is further confirmed by direct phonon calculations. The structurally stable phases also satisfy the thermodynamic stability and mechanical stability criteria. New phases possess extraordinary mechanical properties, similar to diamond, and show insulating bandgap ranging from optimal, 1.45 eV (optimal) to 5.5 eV (large gap). The structural, electronic, and optical properties of $Pm(1) - C_4N_3$ , discussed in detail, indicate possible application in optoelectronic and photovoltaic technologies. The way metric is design, it can be used to predict stability of any material with three-dimensional bonding network. The metric was also able to predict structural stability of other nitride polymorphs with 100% accuracy. We believe that the proposed '*Metric*' will accelerate the search of unknown and unexplored inorganic compounds by quick filtering of dynamically stable phases without expensive density-functional theory calculations.

## Introduction

Compounds made of light elements such as boron, carbon and nitrogen [**1**,**2**] are known for extreme hardness, oxidation resistance and chemical inertness [**3-5**]. These materials find many uses in energy and sustainability applications such as visible-UV light harvesting and photocatalysis [**6-8**], fuel cell and electrolyzer catalyst support [**9**], redox catalysts [**10**,**11**], and some other emerging areas [**12-14**]. Therefore, search for new materials with improved thermodynamic, electronic, and mechanical properties remain a prime challenge for materials science community.

Recently, carbon based inorganic compounds have gained much momentum due to their extraordinary mechanical properties [**4,5**]. In spite of useful physical properties, carbon-nitrides suffer from structural and dynamical instability (unstable phonons), e.g., large pool of known phases are stable only at externally applied pressures [**15**]. The development of new materials requires addressing and resolving fundamental questions concerning their chemical and structural nature in relation to their properties so that they can be designed and optimized for current and future applications. On the one hand, experiments face many challenges related to metastability, selection of precursors, determination of chemical compositions and their crystal structures, whereas calculations are limited to certain stoichiometries [**16**]. Both theoretical (computational) and experimental approaches face significant challenges to develop effective guidelines for the search of structurally stable phases with optimal properties [**16**]. However, determining dynamical stability of newly predicted inorganic compounds from first-principles theory is an expensive affair in terms of computational cost (effective CPU hours).

In this work, we proposed a metric 'M' based on local coordination environment of the constituent elemental sites that removes an existing time-consuming roadblock to assess structural stability, i.e., phonon calculations. The identification of effective local bonding environment is the critical requirement. The charge distribution on local substructures (tetrahedral or trigonal) relative to crystallographic-points (Wyckoff-positions) was calculated for several dynamically (un)stable phases of carbon/boron/silicon-nitride polymorphs using first-principles density-functional theory (DFT). A strong correlation based on local structural motifs with structural stability has been found, which is reflected in the proposed metric 'M'. The metric predicts the stability of all the known and newly predicted phases with 100% accuracy. We believe that the metric will accelerate filtering of useful inorganic compounds from large pool of possible phases, which will save excessive computational cost (CPU hours) needed for phonon calculations.

**Computational Methods**

*Relaxation and electronic properties*: We use first-principles density functional theory (DFT) as implemented in Vienna *Ab-initio* Simulation Package (VASP) [**17**,**18**]. The Perdew, Burke and Ernzerhof (PBE) generalized gradient approximation [**19**] is used with a cut-off energy of 533 eV. The Γ-centered Monkhorst-Pack [**20**] k-mesh was used for Brillouin zone integration during structural-optimization and charge self-consistency. We relaxed each crystal structure with higher convergence of energy ($10^{-6}$ eV) and forces ($10^{-6}$ eV/Å). The energy convergence for each case was checked with respect to k-mesh and energy cut-off. Based on that, we set up high enough energy cutoff. The structural files for known systems were taken from materials project [21]. For band gap calculation, we used hybrid Heyd-Scuseria-Ernzerhof hybrid (HSE06) functional, which includes 25% short-range Hartree-Fock exchange and a range-separation parameter ($\omega$) of 0.11 $Bohr^{-1}$ [22]. We listed both the PBE and HSE06 calculated band gap of newly discovered phases, and for one of the cases, $Pm(1) - C_4N_3$, we show that HSE06 calculated band gap shows good agreement with the more computationally demanding $GW$ method [23,24].

*Formation energy*: The formation energy ($E_{form}$) of new $C_4N_3$ phases was calculated using $E_{form} = E_{total}^{C_4N_3} - \sum_i n_i E_i$, here $E_{total}^{C_4N_3}$ is the total energy of the $C_4N_3$ phase, $n_i$ is the number of C/N atoms in the unit cell, and $E_i$ is the elemental energy of C and N. The $E_{form}$ per atom was calculated by dividing by number of atoms per $C_4N_3$ unit cell.

*Optical properties*: We used self-consistent $GW$ method as implemented within VASP package to calculate quasi-particle energies [17,18]. The random phase approximation (RPA) and Bethe-Salpeter equation (BSE) approaches [25] were used to calculate optical absorption spectra of $Pm\ (1) - C_4N_3$ with a 6×12×6 k-mesh for Brillouin zone integration. The tetrahedron method was employed for Brillouin zone integration [26]. The Gaussian broadening (0.05 eV) was used for BSE calculations.

*Phonon calculation:* The forces on atoms, in general, are zero by symmetry for systems with higher symmetry. However, going down the line, we will notice that our systems of interest are mostly low symmetry, i.e., forces on atoms are not zero by symmetry. Therefore, we must relax the unit cell in order to minimize the force on each atom to 'zero' or as small as possible. This is the one reason, we set very high convergence criteria for energy and forces during the relaxation. This is also important because phonon properties of a material are very sensitive to the forces on atom and cell lattices.

*(a) Density-functional perturbation theory*: Density-functional perturbation theory (DFPT) was used to construct force constant matrix needed to calculate lattice dynamical properties [27-30]. The phonon dispersion plots are done along the high-symmetry direction of the Brillouin zone of crystal structure [31].

*(b) Quasi-harmonic approximation*: We applied quasi-harmonic approximation as implemented in Phonopy with the finite displacement method with a displacement of 0.02 Å [28,31] to calculate harmonic frequencies and normalized eigenvectors. Afterwards, the vibrational contribution of free energies (Helmholtz, entropy, specific heat), Gibbs free-energy, lattice thermal expansion, temperature dependence of (volume, bulk-moduli, Gruneisnen parameter) of one of the newly discovered C-N phases, i.e., $Pm\ (1) - C_4N_3$ was calculated. To estimate above mentioned quantities, we used 11 energy vs volume data points (0.95$V_0$ to 1.05$V_0$), where each data point was further relaxed at a constant volume. All calculations were performed at the Γ-point. The evolution of the cell volume is very common assumption for non-cubic systems within quasi-harmonic approximation too, where previous results are found to be in good agreement, therefore justified [32-35].

## Results and discussion

In spite of great future prospect of inorganic compounds, no design criteria have been presented for semiconductor materials. Most predictions are based on expensive DFT calculations, which makes it difficult to quickly assess the stability of candidate phases, particularly when trying to screen for potential dynamic instabilities.

**Metric definition**: We developed a metric based on local coordination as a surrogate model of charge imbalance. The "*subset of local structure order parameters"* [**36**] was employed to formulate a weight ratio between tetragonal and trigonally bonded systems to exemplify our approach [**36-42**]. The proposed metric uses weights of local substructures (tetragonal/trigonal). Our methodology uses robust neighbor finding approach for measuring maximum resemblance to a target-motif. The metric formulation is based on the assumption that increased covalent nature of bonding of local substructure increases the stability. Therefore, metric ratio (M= tetrahedral (tet)/trigonal (trig)) tends to increase as the local tetrahedral environment is optimized in the crystal structure.

The metric is defined by the average coordination ratio on tetrahedral (tet) and trigonal (tri) substructures with a normalization factor adjusted for element dominant compositions. The metric analyzes bonding character of atomic substructures formed by constituent elements. The metric 'M' sets the ratio for structural stability of inorganic compound for carbon rich compositions as

$$M = \frac{W_{tet}\left({}^{n_2}/{}_{n_1}\right)}{W_{trig}} > 0.6, \tag{1}$$

and the metric 'M' for nitrogen rich composition as

$$M = \frac{W_{tet}\left({}^{n_1}/{}_{n_2}\right)}{W_{trig}} > 0.6, \tag{2}$$

here, the $W_{tet}$ is the average tetrahedral coordination weight of carbon, $W_{trig}$ is the average trigonal coordination weight of nitrogen. The $n_1$,$n_2$ are the number of carbon and nitrogen atoms in the unit cell, respectively. The metric 'M' uses structural features that are sensitive to charge-imbalance at the local sites––where the network of the substructures is the building block of covalently bonded crystalline compounds––as a way to determine whether a given crystalline arrangement is dynamically stable. Therefore, the method can easily be generalized to other systems by choosing the appropriate *subset of local structure order parameters* representing the preferred atomic bonding in the system under investigation.

For example, if we have carbon rich system, the trigonal bonded nitrogen tends to distort itself to accommodate tetrahedral dominant environment for carbon. Therefore, to remove the superficial dominance in weights, we have defined the weighted average in terms of composition in both Eq. (1) & (2). We show later in the metric performance evaluation that this balance mechanism is very critical in achieving increased accuracy of our prediction.

To exemplify, we chose carbon/boron/silicon-based nitride compounds, where preference of covalent type bonding forms desired local bonding substructures. Therefore, nitrides provide fertile ground for testing the metric ‘M’. Later, we show that the metric accurately predicts structural stability based purely on an average of local geometry of each atom type for given crystal types.

**Metric performance**: The nitride polymorphs in the literature can broadly be arranged in three major categories – (a) 3-dimensional (3D) networked, (b) layered (2D), and (c) chains (1D). We have already pointed out in the metric-performance section that the metric works in system that have 3D bonding network, i.e., tetrahedral (tet) vs trigonal (tri)), as discussed in the metric definition section. Therefore, any structures with 2D/1D structural arrangements were eliminated from consideration, as possible prototype precursors in the search for stable networked nitride polymorphs.

For example, the materials project database currently indexes 43 bulk carbon-based nitride phases predicted from DFT. The C-N in those phases can broadly be arranged in three major categories – (a) 3-dimensional (3D) networked (20), (b) layered (13), and (c) chain (10) with average formation energies of 0.906, 0.453 and 2.428 (eV/atom), respectively, above the convex hull. We found a general trend in **Fig. 1** for $C_3N_4$ polymorphs closest to the convex hull that resist 3D C-N network formation. The $C_3N_4$ polymorphs near the convex hull are composed of layered atomic sheets with densities below 2.5 gm/cc. Therefore, these structures were eliminated from consideration, along with one dimensional structure, as possible prototype precursors in the search for stable networked carbon-nitride phases.

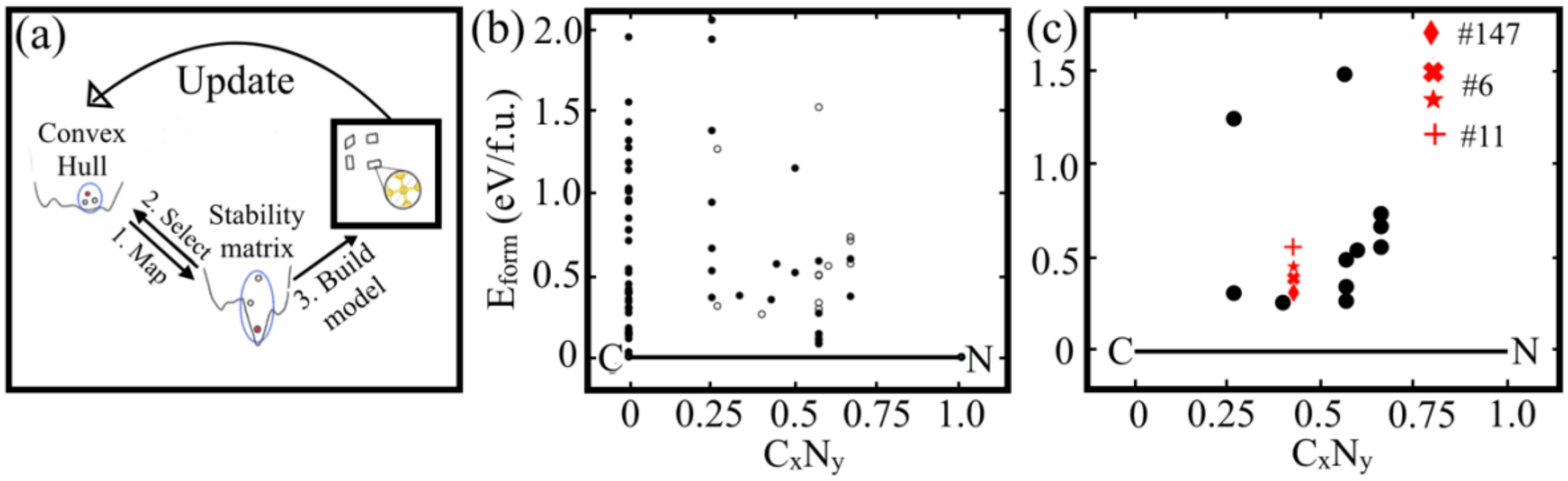

**Figure 1.** (a) Schematic of the flowchart for selecting new matric ‘M’. (b) Convex hull for carbon-nitride phases extracted from materials-project database. The open circles represent the phases used by the model to test the metric (also shown in **Table 1**). (c) Four new phases (#6(1), #6(2), #11, #147) predicted by the proposed metric. Formation

energies in (b) and (c) are calculated using gas phase corrected energies for $N_2$, which is taken from materials project database [**21**].

Rather than relying on random crystal prototyping, we opted to seed our search by altering the composition of the $\beta - C_4N_3$ structure by carbon substitution. In this way, we could compare changes in stability of new phase and attribute them directly to changes in crystal symmetry with emphasis on identifying stable substructures within the modified crystal structure. The C/N coordination weights have been extracted from the existing materials-project database, which has been largely ignored in theoretical predictions and marks a clear gap in the material search space. Regardless of formulaic simplicity, the good predictive performance of the metric stems from a couple of key physical observations:

(1) *Bond angle distortions*: Distortions in bond-angles are easily retained by local substructures. This shifts the average coordination weights away from the preferred valence interaction geometry by forming strong covalent network.

(2) *Maxima in bond angle distortion* either originates or aggregates at symmetry points with lower degrees of freedom. Bader analysis easily captures these charge imbalances.

The aggregate effect of local distortions seems to destabilize the material beyond the harmonic phonon approximation regime.

**Table 1** The proposed metric (M) for structural stability based on charge-ratio. Here, M is the ratio of weights in tetrahedron (tet) vs trigonal (trig) arrangement of C/B/Si and N in carbon-nitride (12), boron-nitride (20), and silicon-nitride (15) polymorphs. We have included phonon plots for the representative B-N and Si-N systems in Fig. 2.

| No. | System | Space group (#) | Metric [M > 0.6] | Structural Stability | |
|---|---|---|---|---|---|
| | | | | Predicted | Calculated |
| **Carbon nitride polymorphs** | | | | | |
| 1 | $CN_2$ | 119 | 0.751 | Yes | Yes |
| 2 | $(CN_2)_2$ | 36 | 0.623 | Yes | Yes |
| 3 | $(CN_2)_4$ | 122 | 0.611 | Yes | Yes |
| 4 | $C_3N$ | 225 | divergent | No | No |
| 5 | $C_3N_4$ | 215 | 2.467 | Yes | Yes |
| 6 | $(C_3N_4)_2$ | 159 | 0.957 | Yes | Yes |
| 7 | $(C_3N_4)_2$ | 176 | 0.878 | Yes | Yes |
| 8 | $(C_3N_4)_2$ | 227 | 0.536 | No | No |
| 9 | $(C_3N_4)_2$ | 220 | 1.109 | Yes | Yes |
| 10 | $C_{11}N_4$ | 16 | 0.597 | No | No |
| 11 | $C_{11}N_4$ | 111 | 0.931 | Yes | Yes |
| **12** | **$(C_3N_2)_4$** | 221 | 0.594 | No | No |
| **13** | **$(C_4N_3)_4$** | 8 | 0.271 | No | No |
| **14** | **$(C_4N_3)_4$** | 10 | 0.368 | No | No |
| **15** | **$(C_4N_3)_4$** | 8 | 0.495 | No | No |
| **Boron nitride polymorphs** | | | | | |
| **1** | **$B_2N_2$** | 194 (1)* | 0.000 | No | No |
| **2** | **$B_2N_2$** | 194 (2)* | 0.000 | No | No |
| **3** | **$B_2N_2$** | 194 (3)* | 0.000 | No | No |
| **4** | **$B_2N_2$** | 194 (4)* | divergent | No | No |
| **5** | **$(B_2N_2)_2$** | 14 | 0.025 | No | No |
| **6** | **$B_2N_2$** | 187 | 0.000 | No | No |
| **7** | **$B_2N_2$** | 186 | 169.1674 | Yes | Yes |
| **8** | **$(B_2N_2)_2$** | 62 | 7.0038 | Yes | Yes |
| **9** | **$B_3N_4$** | 8 | 0.2036 | No | No |
| **10** | **$(B_2N_2)_2$** | 9 | 0.0493 | No | No |
| **11** | **BN** | 216 | 0.9834 | Yes | YES |
| **12** | **$B_3N_3$** | 42 | 2.6961 | Yes | Yes |
| **13** | **$B_9N$** | 139 (1)+ | divergent | No | No |
| **14** | **BN** | 139 (2)+ | divergent | No | No |
| **15** | **$(BN)_{50}$** | 1 (1)# | 0.143 | No | No |
| **16** | **$(BN)_{50}$** | 1 (2)# | 0.0633 | No | No |
| **17** | **$(BN)_{50}$** | 1 (3)# | 0.155 | No | No |
| **18** | **$(BN)_{50}$** | 1 (4)# | 0.0955 | No | No |
| **19** | **$(BN)_4$** | 136 | 0.7435 | Yes | Yes |
| **20** | **$B_{13}N_2$** | 160 | divergent | No | No |
| **Silicon nitride polymorphs** | | | | | |
| **1** | **$Si_{18}N_{24}$** | 159 (1) | 0.492 | No | No |
| **2** | $Si_{12}N_{16}$ | 159 (2) | 0.879 | Yes | Yes |
| **3** | $Si_{12}N_{16}$ | 159 (3) | 0.869 | Yes | Yes |
| **4** | **$Si_3N_4$** | 176 (1) | 1.177 | Yes | Yes |
| **5** | **$Si_{12}N_{16}$** | 62 | 3.469 | Yes | Yes |
| **6** | **$Si_6N_8$** | 176 (2) | 0.828 | Yes | Yes |
| **7** | $Si_6N_8$ | 227 | 0.684 | Yes | Yes |
| **8** | $Si_{12}N_8$ | 221 | 0.442 | No | No |
| **9** | $Si_6N_8$ | 220 | 1.114 | Yes | Yes |
| **10** | $Si_{45}N_{60}$ | 1 (1)** | 0.896 | Yes | Yes |
| **11** | $Si_{45}N_{60}$ | 1 (2)** | 0.995 | Yes | Yes |
| **12** | $Si_6N_8$ | 194 | 0.000 | No | No |

** All 194-$B_2N_2$ are structurally unstable, so we provide only representative phonon.*
*+ Both 139 (1) and 139 (2) are structurally unstable, so we provide only representative phonon plot.*
*# All P1 BN phases are isostructural and structurally unstable, so we provide only representative phonon plot.*
*** P1-$Si_{45}N_{60}$ same structures both in number of atoms and cell dimensions.*

*Phonon dispersion of Carbon-, Boron-, and Silicon-nitride Polymorphs:* We employed *matminer featurizer* to extract the coordination character and weights of nitride polymorphs for given structures in materials project database [**21**]. We list the net atomic charge at neighboring (X (C, B, Si) or Y (C, N)) sites from Bader analysis at Wyckoff symmetry point in **Table 1**. The metric analyzes bonding character of atomic substructures formed by constituent elements in inorganic compounds. To verify the accuracy of our prediction and investigate the ground truth of structural stability, in Fig. 2, we plot phonon dispersion at zero pressure. The metric does not satisfy structural stability criteria for (a) C-N: #227, #11, #211, #215, and #225, (b) B-N: #194, #14, and #187, and (c) Si-N: #159. This is also reflected in unstable phonons, i.e., negative modes in **Fig. 2**. Notably, the metric in Eq. (1) & (2) correctly predicts the structural stability as shown in Fig. 2 and Table 1. Just to avoid any confusion, the metric M [x] can be understood as a binary function (true/false), i.e., M [$x \geq 0.6$] = 1 (true; stable); and M[$x<0.6$] = 0 (false; unstable). Here, the metric is not intended to be used to predict the strength of the phonon modes as the strength of phonon modes is mainly related either to first order (discontinuous change of crystal structure change) or second order (infinitesimal atomic displacement and continuous change of crystal structures) phases transition.

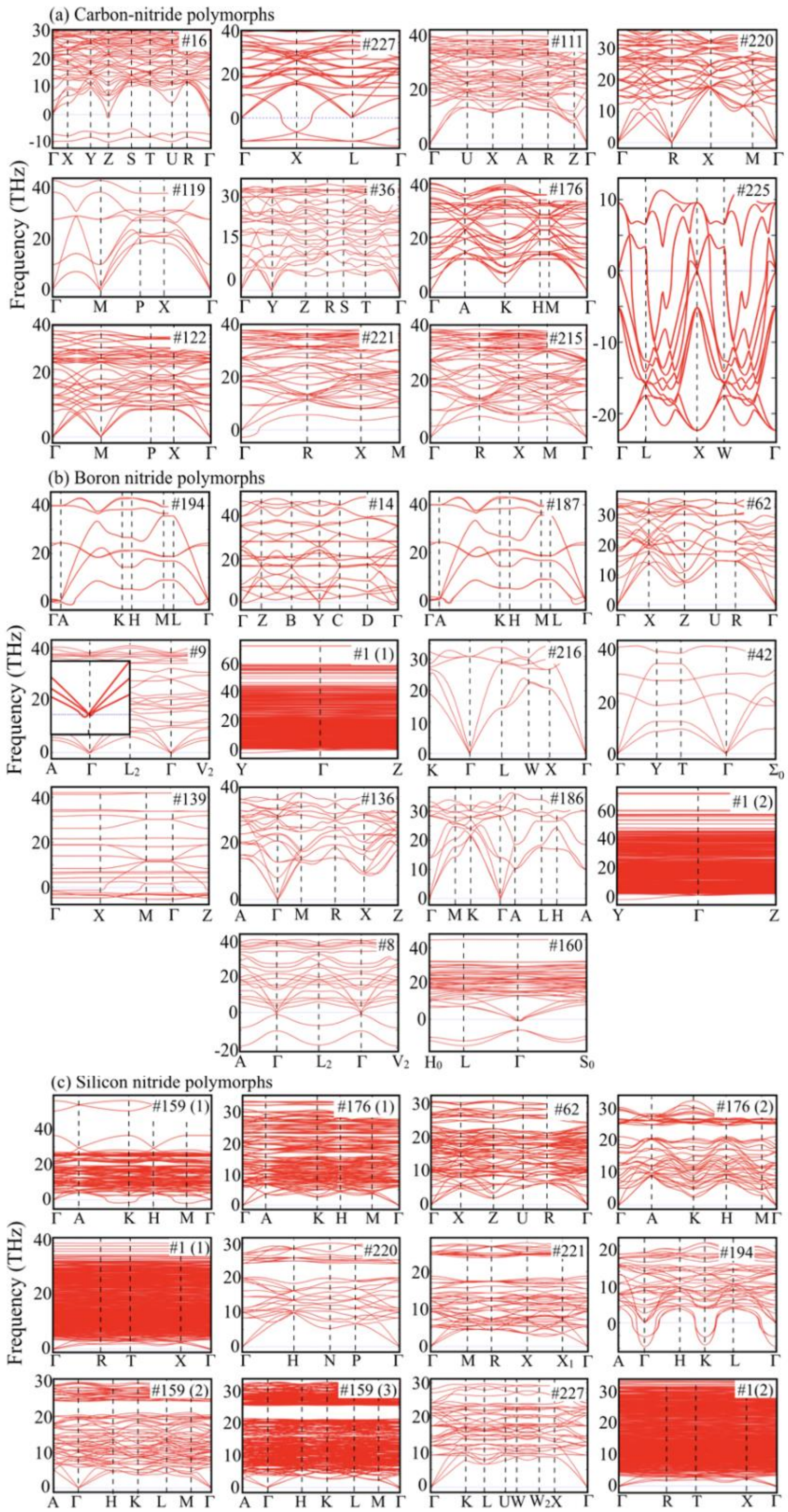

**Figure 2**. DFT calculated phonons showing stability for (a) carbon-nitride (15), (b) boron-nitride (20), and (c) silicon-nitride (12) polymorphs as predicted using metric 'M' in Table 1.

*Stability (structural) analysis of* $C_{11}N_4$: We attempt to investigate and establish the possible reasoning for structural stability based on charge imbalance and metric score. We choose $C_{11}N_4$ as an example to discuss the relevance of the proposed metric. 'C' and 'N' in $C_{11}N_4$ form two local substructures with tetrahedron and trigonal bonding environment. The metric based on simple charge imbalance analysis on local sub-structure formed by C-C, C-N, and N-N network in $C_{11}N_4$ predicts space group #16 as structurally unstable while structural stability for #111, also highlighted in **Table 1**. The carbon in space group #16 in **Fig. 3a** is positioned at the special symmetry point (0,0,0) with $+0.21$ charge, also highlighted in **Table 1**. The C at (0,0,0) for space group #16 is a fixed symmetry point with no possible shift for C using translation symmetry to accommodate extra strain in bond-angles due to excess charges. Here, the Bader charge imbalance coincides with maximal distortion of the ideal tetrahedral coordination. The tetrahedral weight feature of C (0,0,0) has the lowest metric score in **Table 2**, i.e., M=0.273. Notably, the special attention should be given to carbon in space group #111 as shown in **Fig. 3b** located at the special Wyckoff points with lowest symmetry, i.e., (0, 0, 0) and (½ ½ 0), see **Table 2**. The C at (0,0,0) and (½ ½ 0) in space group #111 are very close to charge neutral with charges $+0.01$ and $-0.07$, respectively. This leads to higher metric scores of 0.955 and 0.949, which is indication of structural stability based on proposed metric in Eq. (1). This is also a crucial observation as it directly confirms a possible correlation between ideal coordination environment and dynamical stability of carbon nitride polymorphs. The rank column in Table 2 is the trace of the matrix that also represents the basis of the Wyckoff position, which can also be understood as the multiplicity of the Wyckoff points. If Wyckoff point is a special symmetry point, i.e., (0, 0, 0) or (½ ½ ½), then rank is 0. However, if Wyckoff point is (x,y,0) or (x,y,z) then the rank will be 2 or 3, respectively.

**Table 2**. The charge on local tetragonal (C) and trigonal (N) substructures of $P\bar{4}2m{-}C_{11}N_4$ (#16) and $P222{-}C_{11}N_4$ (#111) was analyzed using Bader charge analysis. The metric score for $P\bar{4}2m{-}C_{11}N_4$ (#16) and $P222{-}C_{11}N_4$ (#111) using **Eq. 1** is 0.597 and 0.931, respectively.

| Element | Ideal coordination | Weight | Charge | Wyckoff positions | Rank |
|---|---|---|---|---|---|
| $P\bar{4}2m{-}C_{11}N_4$ (#16) | | | | | |
| **C** | tetrahedral | 0.273 | 0.21 | 0, 0, 0 | 0 |
| **C** | tetrahedral | 0.940 | 1.19 | 0, ½, ½ | 0 |
| **C** | tetrahedral | 0.643 | 0.62 | 0, 0, z | 1 |
| **C** | tetrahedral | 0.374 | -0.11 | x, y, z | 3 |
| **C** | tetrahedral | 0.272 | -0.02 | 0, ½, 0 | 0 |
| **C** | tetrahedral | 0.896 | 0.54 | ½, ½, z | 1 |
| **N** | trigonal | 0.337 | -0.82 | x, y, z | 3 |
| $P222{-}C_{11}N_4$ (#111) | | | | | |
| **C** | tetrahedral | 0.955 | -0.01 | 0, 0, 0 | 0 |
| **C** | tetrahedral | 0.949 | 0.07 | ½, ½, 0 | 0 |
| **C** | tetrahedral | 0.936 | 0.59 | 0, ½, z | 1 |
| **C** | tetrahedral | 0.930 | -0.02 | x, x, z | 2 |
| **C** | tetrahedral | 0.950 | 1.24 | 0, 0, ½ | 0 |
| **N** | trigonal | 0.368 | -0.88 | x, x, z | 2 |

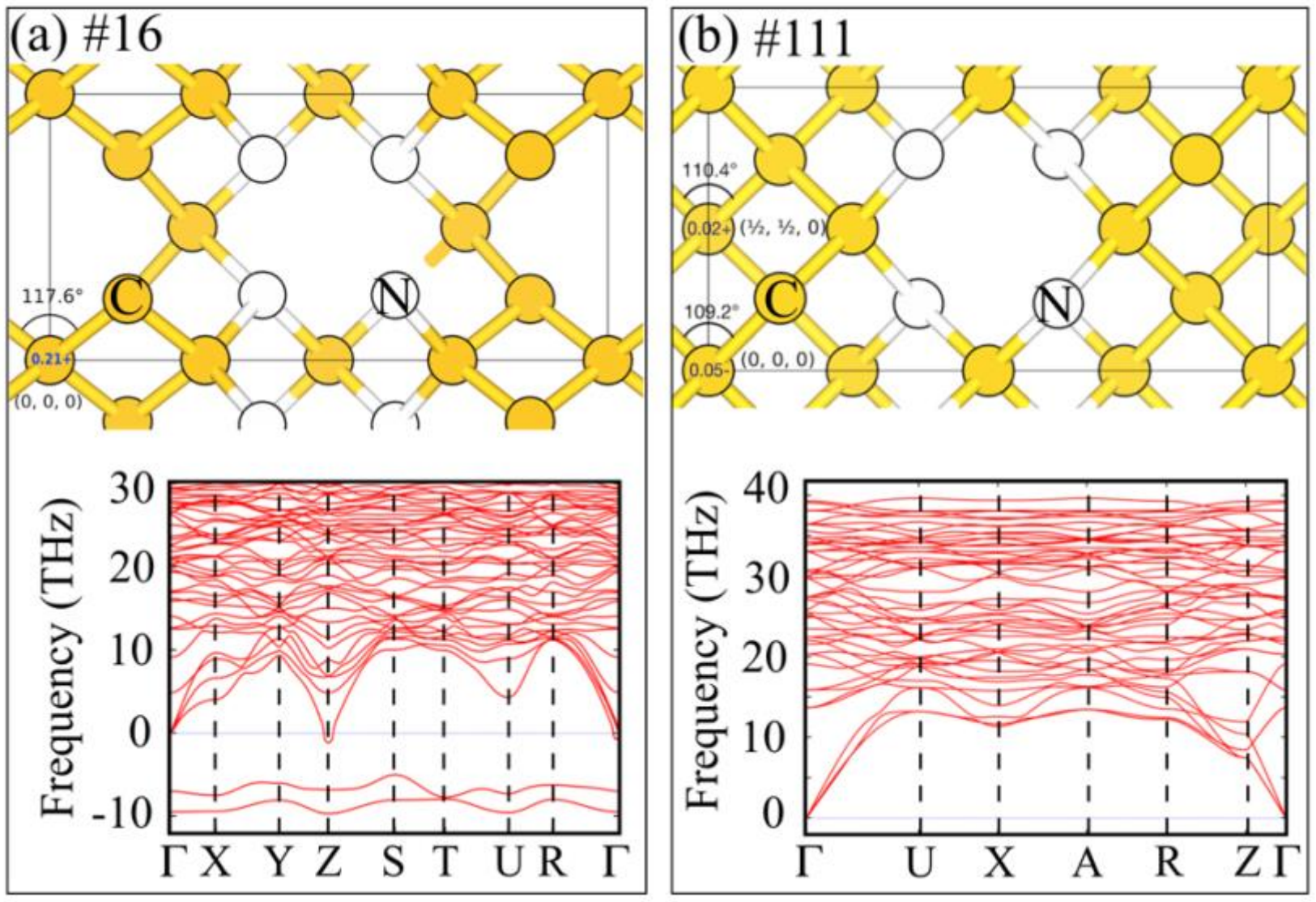

**Figure 3.** The crystal structure and phonon dispersion for $C_{11}N_4$ polymorphs with space group (a) #16, and (b) #111 are shown.

Based on overall analysis of charge imbalance, metric scores for $P\bar{4}2m{-}C_{11}N_4$ *and* $P222{-}C_{11}N_4$ are 0.597 and 0.931, respectively. The phonon dispersion for $C_{11}N_4$ with #16 and #111 space group is shown in **Fig.3a-3b**. Two imaginary modes for space group #16 in phonon dispersion along the Brillouin zone in **Fig. 3a** and soft modes at Γ suggests structural instability, whereas no imaginary eigen modes in phonon dispersion for space group #111 in **Fig. 3b** at zero pressure shows structural stability. The structural stability predicted using metric 'M' based on simple charge analysis conforms with the direct phonon calculations, which further confirm our predictions.

**Metric prediction – new carbon nitride phases**

The qualitative nature of the metric `M' in **Eq. (1) & (2)** is based on local structural information and charges, which was tested for 15 carbon-nitride, 4 boron-nitride, and 5 silicon-nitride polymorphs as shown in **Table 1**. In this work, however, we focus on the extended carbon-nitride ($\mathrm{C_xN_y}$) phase space to identify possible structurally stable phases as it has been at the center of extensive research and has large number of available test cases. The metric 'M' identifies four new carbon nitride phases with 4 (C):3(N) stoichiometry as marked by red in **Fig. 1c**. To establish the reliability of metric, the metric-predicted structural stability of newly discovered phases is compared with direct DFT calculations (phonons) and is tabulated in **Table 3**. We also plot the phonon dispersion in **Fig. 4,** which accurately matches with the structural (in)stability predictions from the metric. The good agreement between predictions and direct calculation is very exciting given that new set of compounds have been identified by the simple screening.

**Table 3.** Metric 'M' predicted structurally stable new carbon-nitride phases, with $C_4N_3$ stoichiometry. Phases are arranged in order of increasing `M'.

| C-N Systems (z=multiplicity) | Space group (#) | Stability metric [M > 0.6] | Structural Stability | |
|---|---|---|---|---|
| | | | **Predicted** | **Calculated** |
| **$C_4N_3$ (z=2)** | $Pm$ (#6) | 0.664 | Yes | Yes |
| **$C_4N_3$ (z=4)** | $Pm$ (#6) | 0.968 | Yes | Yes |
| **$C_4N_3$ (z=4)** | $P\bar{3}$ (#147) | 0.843 | Yes | Yes |
| **$C_4N_3$ (z=2)** | $P2_1/m$ (#11) | 0.289 | No | No |

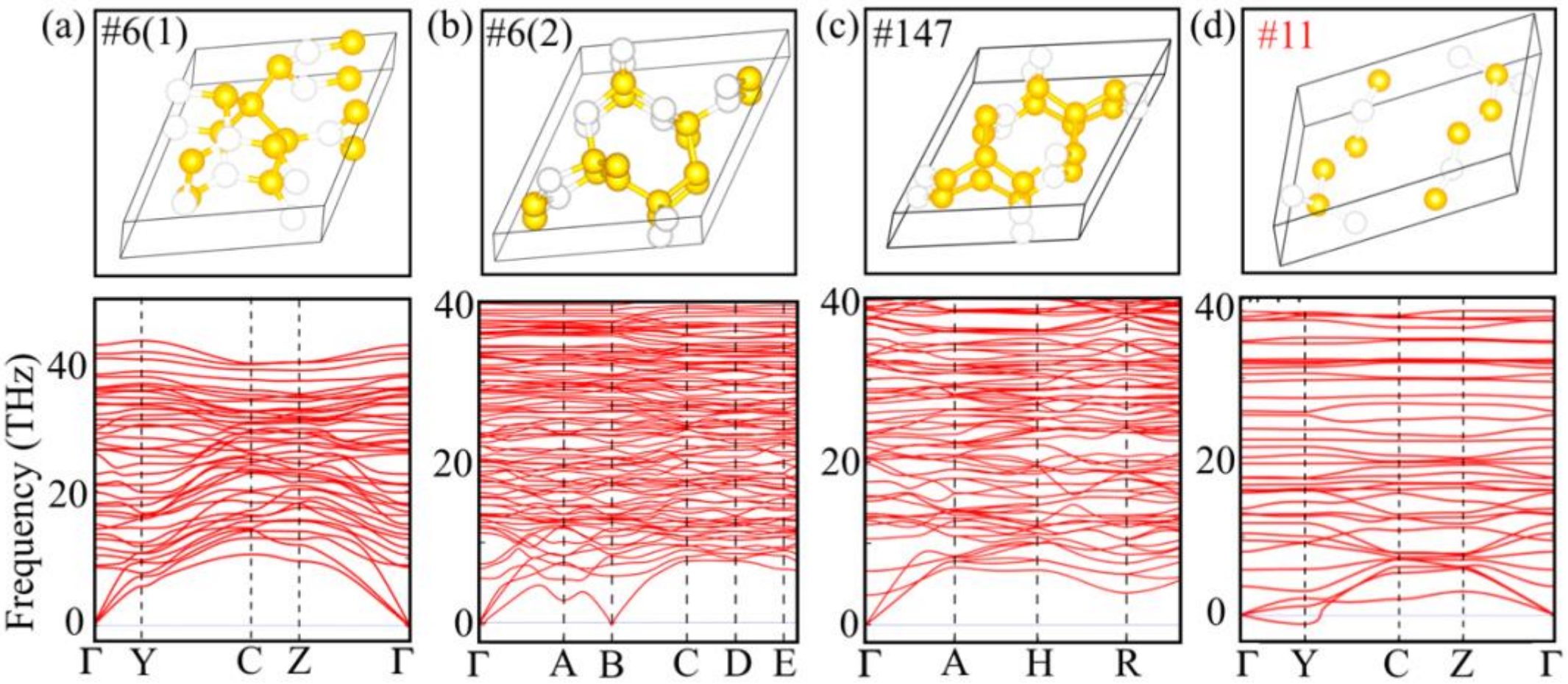

**Figure 4.** Phonon dispersion for $C_4N_3$ with (a) $Pm$ (1) (#6), (b) $Pm$ (2) (#6), (c) $P\bar{3}$ (#147), and (d) $P2_1/m$, (#11) space group confirms our predictions in Table 2.

*Space group analysis of new phases*: The $\beta - C_4N_3$ phase belongs to the space group $P6_3/m$, which has two unique N atoms in the asymmetric unit cell. One is located in mirror plane at point (x, y, ¼) and the other at a roto-inversion point (1/3,2/3,1/4). A prototype $C_4N_3$ structure can be generated by substitution at either of the two Wyckoff positions 6$h$, or 2$c$ of $\beta - C_4N_3$, respectively. Two possible crystal symmetries can emerge by carbon substitution: (a) Pm symmetry: by substitution on a mirror plane; and (b) $P6_3/m$ by substitution at the roto-inversion point with site symmetry '$6 -$'. The carbon substitution on a mirror plane gives $Pm$ (1)$-C_4N_3$ phase, which is a novel structure with C atoms regularly arranged around the mirror plane. The basic unit cell consists of 14 C (8) and N (6) atoms conform with the $Pm$ symmetry. Although C-substitution at roto-inversion point (1/3,2/3,1/4) of $P6_3/m$ does not affect space group but a carbon atom placed in that position is unstable towards displacement along the c-axis. However, if not displaced, the C coordination would form an unlikely substructure, which wouldn't satisfy the proposed metric 'M'. C in tetrahedron sub-structure in $Pm$ (1)$-C_4N_3$ becomes positively charged by losing electrons, whereas the trigonal substructure gains some and becomes negatively charged, see **Table 5**. This charge imbalance created by switching C/N in $\beta - C_4N_3$ stabilizes the new $Pm\ (1){-}C_4N_3$. This is also reflected in the metric M=0.664 in **Table 4**.

**Table 4**. The charge analysis on local substructures, i.e., tetragonal-C and trigonal-N substructures of $Pm\ (1){-}C_4N_3$. The metric 'M' is defined by sum average of weights on tetrahedrally (tet) arranged 'C', multiplied by (8/6) and divided by sum average on trigonally (trig) arranged 'N' (see eq. 1). 'M' is 0.664 for $Pm\ (1){-}C_4N_3$ [**36**].

| Elements | Wyckoff Position | Ideal coordination | Charge | Weight | Rank |
|---|---|---|---|---|---|
| **C** | x, 0, z | tetrahedral | 1.64 | 0.722 | 2 |
| **C** | x, 1/2, z | tetrahedral | -0.02 | 0.65 | 2 |
| **C** | x, 1/2, z | tetrahedral | 0.98 | 0.738 | 2 |
| **C** | x, 0, z | tetrahedral | 1.28 | 0.586 | 2 |
| **C** | x, 0, z | tetrahedral | 1.01 | 0.78 | 2 |
| **C** | x, 1/2, z | tetrahedral | 0.98 | 0.855 | 2 |
| **C** | x, 0, z | tetrahedral | 1.98 | 0.756 | 2 |
| **C** | x, 0, z | tetrahedral | 4.0 | 0.0 | 2 |
| **N** | x, 1/2, z | trigonal | -0.94 | 0.482 | 2 |
| **N** | x, 0, z | trigonal | -1.65 | 0.736 | 2 |
| **N** | x, 0, z | trigonal | -2.36 | 0.442 | 2 |
| **N** | x, 0, z | trigonal | -2.3 | 0.546 | 2 |
| **N** | x, 1/2, z | trigonal | -1.59 | 0.749 | 2 |
| **N** | x, 0, z | trigonal | -3.01 | 0.0 | 2 |

The space group $P2_1/m{-}C_4N_3$ (#11) is obtained by fully relaxing of $\beta - C_4N_3$ by switching C and N positions. The negative modes in space group #11 in **Fig. 4d** correspond to carbon displacement at two equivalent roto-inversion points. The negative frequency phonons were followed in deciding C substitution in space group #11. Followed by this, all degrees of freedoms (lattice and ions both) of the new carbon-nitride phase were fully relaxed. Our symmetry analysis shows new space group $P\overline{3}$ (#147) for the $C_4N_3$. The predicted M=0.843 > 0.6 for $Pm\ (2) - C_4N_3$ indicates its structural stability (supplement Section S1 (b)), which is further confirmed by direct phonon calculations in **Fig. 4c**.

Clearly, the proposed metric rests on the assumption that the physical behavior of crystalline compounds is the result of sub-structural motifs, e.g., ratio of tetrahedron/trigonal coordination. The sub-structural motifs can also be understood as a marking or preference of C/N/Si to form tetrahedral co-ordination in nitride polymorphs, which directly correlates with physical properties of inorganic compounds. For example, metric accurately identifies new structurally stable carbon-nitride phases. Therefore, the proposed metric is philosophically similar to other rules that have been discovered in solid-state chemistry, such as by Hume-Rothery. The only difference is that this rule is about dynamic (structural) stability, not thermodynamic stability.

*Structural properties and phase stability*: The newly predicted phases show monoclinic ($Pm\ (1), Pm\ (2), P2_1/m$ ) and hexagonal symmetry ($P\overline{3}$). The $Pm\ (1)$ and $Pm\ (2)$ has same symmetry, however, $Pm\ (2)$ (28 atoms per cell) is twice the volume of $Pm\ (1)$ (14 atoms per cell) phase. We also calculated the formation enthalpy ($E_{form}$) for all four phases as shown in **Table 5**. The $E_{form}$ is a fundamental quantity, where $E_{form} < 0$ indicates the thermodynamics stability of the compound and each phase satisfies the stability criteria. Notably, the proposed metric and direct phonon calculations shows that $P2_1/m$ phase is structurally unstable. This indicates that the phase-stable only criteria can present a deceptive picture. Therefore, the metric presented here can be very useful for design and discovery of new structurally stable materials that will also be more computationally efficient with respect to DFT based methods.

*Band gap*: Our electronic structure study shows that structurally stable phases show optimal band gap ($Pm\ (1) - C_4N_3$ = 1.32 eV) to wide band gap ($Pm\ (2) - C_4N_3$= 5.88 eV and $P\overline{3} - C_4N_3$ =5.77 eV), whereas the structurally unstable phase ($P2_1/m - C_4N_3$) is metallic in nature. To understand the effect of exchange-correlation, we calculated the band gap for of $Pm\ (1) - C_4N_3$ from PBE (GGA), GW, and HSE06 hybrid functionals as shown in **Table 5**. The GGA severely underestimates the band gap (0.30 eV), whereas hybrid functional HSE06 (1.32 eV) and *GW* (1.45 eV) show significant improvement. The 1.30-1.45 eV band gap for $Pm\ (1) - C_4N_3$ predicted from hybrid/GW functionals falls in the visible (light) range, which can be important for photovoltaic or optoelectronic applications.

*Mechanical stability*: The elastic parameters ($C_{ij}$) can describe some fundamental material characteristics such as thermal properties including specific heat, thermal expansion, Debye temperature and Gruneisen parameter, therefore, the discussion of $C_{ij}$'s of new phases is indispensable. The elastic response and mechanical stability of any crystalline phases can be determined from $C_{ij}$ [**43**]. The elastic properties of the four newly discovered phases were investigated using stress-strain approach as implemented in VASP. The stress-strain approach performs six finite distortions to the lattice to derive the elastic constants using both rigid ions as well as allowing relaxation of the ions [**44**]. Four $C_4N_3$ phases can be categorized as (a) monoclinic symmetry group ($Pm$ (1), $Pm$ (2), $P2_1/m$), and (b) hexagonal symmetry group ($P\overline{3}$). The Born criteria for mechanical stability for two symmetry group can be written as [**43**]:

(a) *Monoclinic symmetry group*:

$(I - VI)\ C_{ii} > 0, here\ ii = 11, 22, 33, 44, 55, 66$

$(VII)\ (C_{11} + C_{22} + C_{33}) + 2(C_{12} + C_{13} + C_{23)}) > 0,$

(VIII) $C_{33} \times C_{55} - (C_{35})^2 > 0,$

(IX) $C_{44} \times C_{66} - (C_{46})^2 > 0$

(X) $C_{22} + C_{33} - 2 \times C_{23} > 0,$

(XI) $C_{22}(C_{33} \times C_{55} - (C_{35})^2 + 2(C_{23} \times C_{25} \times C_{35}) - (C_{23})^2 \times C_{55} - (C_{25})^2 \times C_{33} > 0,$

(XII) $2[C_{15} \times C_{25}(C_{33} \times C_{12} - C_{13} \times C_{23}) + C_{15} \times C_{35}(C_{22} \times C_{13} - C_{12} \times C_{23}) + C_{25} \times C_{35}(C_{11} \times C_{23} - C_{12} \times C_{13})]$-[ $(C_{15})^2(C_{22} \times C_{33} - (C_{33})^2) + (C_{25})^2(C_{11} \times C_{33} - (C_{13})^2) + (C_{35})^2(C_{11} \times C_{22} - (C_{12})^2)]$+$C_{55} * g > 0, here\ g = C_{11} \times C_{22} \times C_{33} - C_{11} \times (C_{23})^2 - C_{22} \times (C_{13})^2 - C_{33} \times (C_{12})^2 + 2C_{12} \times C_{13} \times C_{23};$

(b) *Hexagonal symmetry group*:

$(I) C_{11} - C_{12} > 0;$

$(II)\ 2(C_{11})^2 < C_{33}(C_{11} + C_{12});$

$(III)\ C_{44} > 0.$

Mechanical stability test was performed on all four $C_4N_3$ phases, we found that $P2_1/m{-}C_4N_3$ failed the Born criteria (highlighted in red in Table 3). The mechanical instability of $P2_1/m$ phase is in agreement with the structural instability predicted from the proposed Metric "M" in **Table 2** and direct phonon calculations in **Fig. 4d**. We used DFT calculated $C_{ij}$'s to derive other useful quantities such as elastic moduli (Bulk, Shear, and Young's), melting-temperature, Debye temperature, Pugh's ratio (K/G), and Poisson's ratio for the four phases and tabulate them in **Table 5**. All three structurally and mechanically stable phases show remarkably very high elastic strength similar to diamond, and very higher melting temperatures. Notably, the higher strength in stable phases arise from the three-dimensional C-N framework that have much shorter covalent bonds compared to P$2_1/m$. For example, C-N framework is discussed for $Pm$ (1)$-C_4N_3$ in **Fig. 5** and **Fig. S1**.

The bulk modulus (K) of a material is the resistance to external deformation, which also indicates towards the bonding characteristics of the material [**45**]. On the other hand, the shear (G), and Young's moduli (E) describe the resistance to change in shape due to shear and uniaxial tensions (degree of stiffness), respectively [**45**]. Like $C_{ij}$'s, B, G and E of stable phases of $C_4N_3$ ($Pm$ (1), $Pm$ (2), $P\bar{3}$) points towards the similar behavior as shown in Table 2.

The degree of brittleness or ductility is very important to understand for any practical application of a material. Unlike ductility, brittleness makes materials less deformable before fracture. The Pugh's ratio ($k = K/G$) has been used as the limit to determine degree of brittleness or ductility, here, 1.75 and higher $k$ makes the material ductile, otherwise brittle [**46**]. The low $k$ values for stable $C_4N_3$ in **Table 5** suggest that newly discovered phases are brittle in nature. The Poisson ratio (ν) is used as an indication for the degree of covalent nature, i.e., smaller the **ν** stronger is the covalent character [**47**]. Smaller **ν** for three stable phases of $C_4N_3$ which suggest towards covalent bonding nature that gives higher strength.

The secondary properties such as Debye ($T_D$), and melting ($T_m$) temperatures, essential in processing for any industrial application, can be calculated from elastic constants [**48**]. This information is required to estimate the maximal possible operating temperature range. The high $T_D$ and $T_m$ for $C_4N_3$ in **Table 5** stem from strong covalent bonding. High $T_D$ usually leads to high thermal conductivity [**49**]. This can be useful in application as high thermal conductors to help systems dissipate heat, which will delay/stop possible physical breakdown of devices due to excess heat.

**Table 5.** Lattice parameters, formation enthalpy, bandgap, and elastic properties of new *Pm (1)*, $P2_1/m$, *Pm (2)*, and $P\bar{3}$ phases calculated from density-functional theory. Our mechanical stability test confirms that $P2_1/m$ phase is structurally unstable.

| **Space group** | $\boldsymbol{a}$ | $\boldsymbol{b}$ | $\boldsymbol{c}$ | $\boldsymbol{\alpha}$ | $\boldsymbol{\beta}$ | $\boldsymbol{\gamma}$ | $\boldsymbol{E_{form}}$ | $\boldsymbol{E_{Gap}(eV)}$ | | |
|---|---|---|---|---|---|---|---|---|---|---|
| | **(Å)** | | | **(º)** | | | **(eV)** | **PBE** | **GW** | **HSE** |
| $\boldsymbol{Pm}$ **(1)** | 6.275 | 2.509 | 6.448 | 90 | 116.7 | 90 | -1.03 | 0.30 | 1.45 | 1.32 |
| $\boldsymbol{Pm}$ **(2)** | 6.539 | 4.809 | 6.552 | 90 | 119.5 | 90 | -1.13 | 4.22 | -- | 5.88 |
| $\boldsymbol{P\bar{3}}$ | 6.628 | 6.628 | 4.723 | 90 | 90 | 120 | -1.13 | 4.13 | -- | 5.77 |
| $\boldsymbol{P2_1/m}$ | 8.618 | 5.996 | 2.347 | 90 | 90 | 115.2 | -0.88 | -- | -- | -- |
| | | | | | | | | | | |
| $\boldsymbol{C_{ij}}$ **(GPa)→** | $\boldsymbol{C_{11}}$ | $\boldsymbol{C_{22}}$ | $\boldsymbol{C_{33}}$ | $\boldsymbol{C_{44}}$ | $\boldsymbol{C_{55}}$ | $\boldsymbol{C_{66}}$ | $\boldsymbol{C_{12}}$ | $\boldsymbol{C_{13}}$ | | $\boldsymbol{C_{23}}$ |
| $\boldsymbol{Pm}$ **(1)** | 910 | 827 | 907.5 | 153.5 | 246.3 | 236.4 | 52.9 | 77.4 | | 29.8 |
| $\boldsymbol{Pm}$ **(2)** | 617 | 1013 | 711 | 310 | 230 | 289 | 157 | 224 | | 125 |
| $\boldsymbol{P\bar{3}}$ | 657 | 657 | 1114 | 344 | 344 | 228 | 201 | 137 | | 137 |
| $\boldsymbol{P2_1/m}$ | 92 | 170 | 916 | 153 | -92 | 54 | 78 | 15 | | -2 |
| | | | | | | | | | | |
| $\boldsymbol{C_{ij}}$ **derived properties →** | **K** | **G** | **E** | **P-wave** | $\boldsymbol{\upsilon}$ | $\boldsymbol{k}$**=K/G** | | $\mathbf{T_D}$ **(K)** | | $\mathbf{T_m}$ **(K)** |
| | **(GPa)** | | | | | | | | | |
| $\boldsymbol{Pm}$ **(1)** | 329 | 293 | 678 | 720 | 0.157 | 1.125 | | 1598 | | 3652 |
| $\boldsymbol{Pm}$ **(2)** | 372 | 288 | 688 | 757 | 0.192 | 1.292 | | 1625 | | 4039 |
| $\boldsymbol{P\bar{3}}$ | 375 | 313 | 736 | 793 | 0.173 | 1.198 | | 1684 | | 4036 |
| $\boldsymbol{P2_1/m}$ | 151 | 95 | 237 | 279 | 0.239 | 1.583 | | 908 | | 1639 |

### Thermal, electronic, and mechanical properties of $\boldsymbol{Pm}$ $(\mathbf{1})-\boldsymbol{C_4N_3}$ (#6)

*Vibrational contribution to Helmholtz free energy ($F_{vib}$), entropy ($S_{vib}$), and specific heat ($C_v$)*: The thermodynamic properties of $Pm$ $(1)-C_4N_3$ are derived from the phonon spectrum calculated using DFT-QHA method by taking zero-point energy into account. The DFT-QHA analysis provides useful information on vibrational contribution to Helmholtz free-energy ($F_{vib}$), entropy ($S_{vib}$) and specific-heat at constant volume ($C_v$) [**31,40,41**]. For DFT-QHA, we calculate the total energy and phonon free energies at '11'different equilibrium volume points for $Pm$ $(1)-C_4N_3$ as shown as a function of temperature and then fitted in **Fig. 5a** (see related phonon plot in Fig. S2). The free energy contribution decreases monotonously with increasing temperature, while entropy contribution increases with temperature. The specific heat in **Fig. 5a** follows the Debye $T^3$ law in the low temperature region (T < ~250 K), whereas it is close to a fixed value with approximately 21.4 J/K/mol (~25 J/K/mol per atom) in high temperature region (above 250 K). The findings are in good agreement with the Dulong-Petit law [**42**].

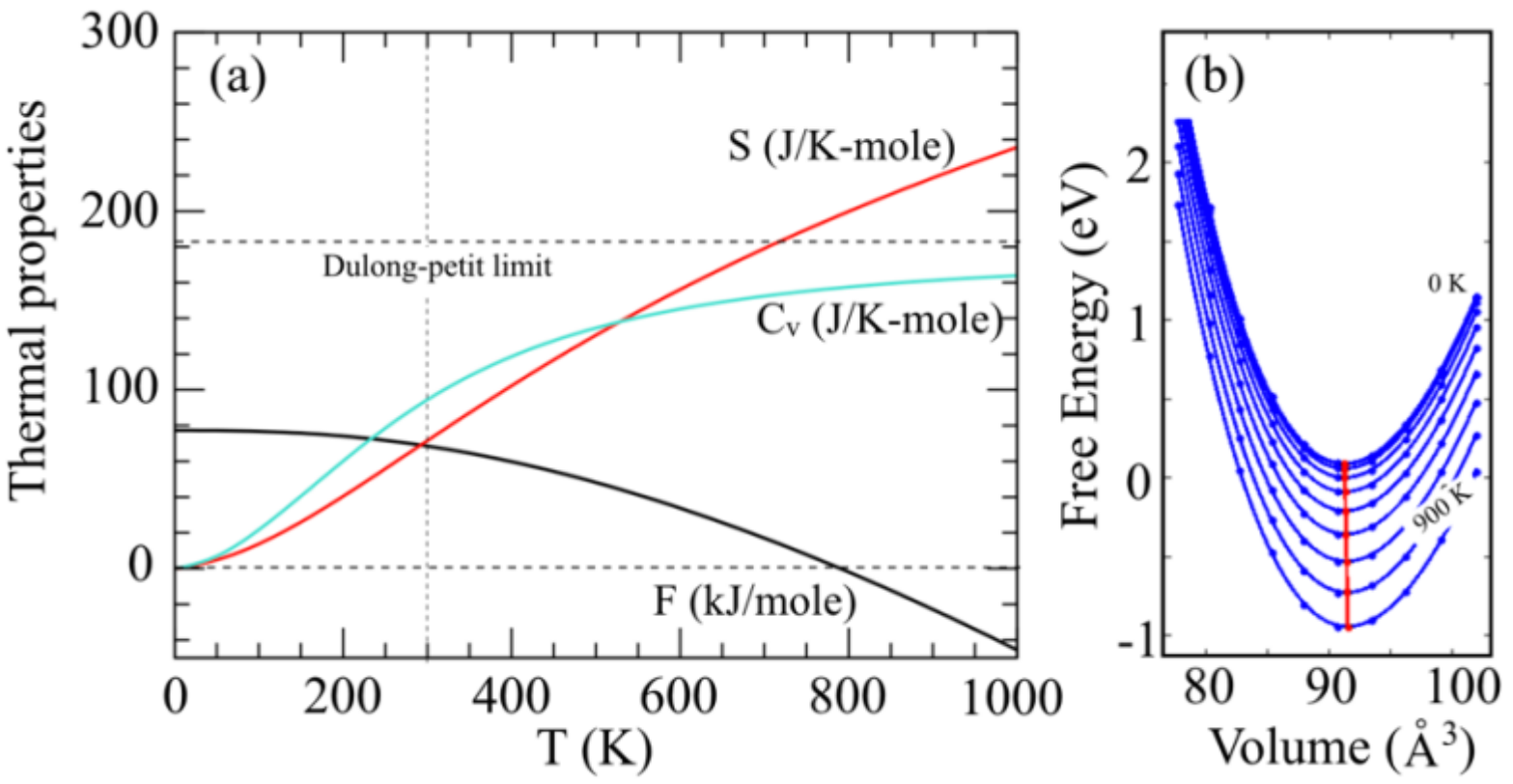

**Figure 5**. We estimate temperature dependence of vibrational contribution of (a) Helmholtz free energy, entropy, and specific heat for $Pm\ (1)\!-\!C_4N_3$ from quasi-harmonic approximation [23,34]. (b) The Helmholtz free energies as a function of volume for $Pm\ (1)\!-\!C_4N_3$ from 0 to 900 K at every 100 K. The equilibrium volumes are marked by red line crossing each free energy curve.

We investigated the free energy dependence on unit cell volume of $Pm\ (1)\!-\!C_4N_3$ between 0 to 900 K at every 100 K as displayed in **Fig. 5b**. The equilibrium volumes are marked by solid red circles along the line are obtained by fitting Vinet equation of state [**32**]. The evolution of the cell volumes is common assumption for non-cubic systems within quasi-harmonic approximation, where previous results are found to be in good agreement [**32-35**], therefore justified.

*Structural property, negative thermal expansion, and Grüneisen parameter*: In **Fig. 6(a)**, volume and bulk moduli are plotted vs temperature. The volume at 0 K is 91.295 Å$^3$, which shrinks (-) by 0.02% to 91.28 Å$^3$ at 130 K then increases by 0.35% to 91.6 Å$^3$ at 1000 K. For 0.02% change in volume at 130 K, we see 0.15% change, i.e., 0.3 GPa, in bulk moduli with respect to 0 K. Maximum change in bulk moduli at 1000 K with respect to 130 K is only 0.87%. We believe that this comes from the strong covalent nature of bonding in $Pm\ (1)\!-\!C_4N_3$ as discussed in mechanical stability section provides higher strength, which makes them hard to break.

Grüneisen parameter in **Fig. 6b** directly reflects the anharmonicity in phonon mode, which is also related to third order force constants [**31**, **51**]. Notably, the Grüneisen parameter is negative between 0 to 130 K, which is the signature of stiffness of transverse acoustic modes [**52**]. Volumetric thermal expansion coefficient $\left(\beta = \frac{1}{V}\left[\frac{\partial V(T)}{\partial T}\right]\right)$ in **Fig. 6c** was be obtained from the calculated equilibrium volumes [**53**]. The negative thermal expansion (NTE) was observed from 0 K to130 K, which is on average -0.49×10$^{-6}$ K$^{-1}$, whereas positive thermal expansion was observed for 130 K and above with average thermal coefficient of +5.34×10$^{-6}$ K$^{-1}$. The NTE in **Fig. 6c** closely relates to the anharmonicity of lattice vibrations [**54**]. The negative modes in Grüneisen parameter serves as a bridge between the behavior of the lattice vibration and the NTE for $Pm$ $(1){-}C_4N_3$ [**55**].

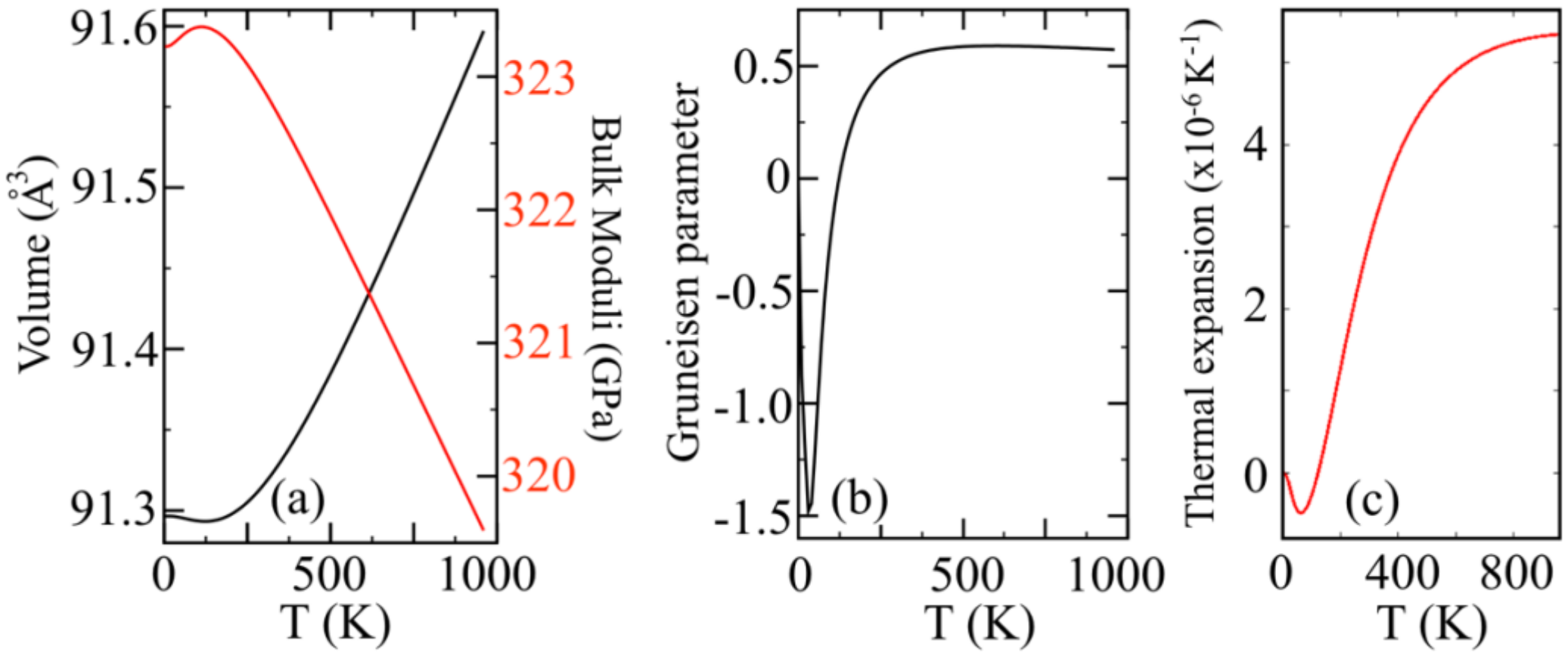

**Figure 6.** (a) Volume/bulk-moduli vs temperature, (b) Grüneisen parameter, and (c) volumetric thermal expansion coefficient for $Pm$ $(1){-}C_4N_3$.

*Electronic-structure properties:* Having established promising compositions and their candidate structures, we provide quantitative analysis for one of the predicted phases, i.e., $Pm$ $(1){-}C_4N_3$, where C occupies the mirror plane. The optimized $Pm$ $(1){-}C_4N_3$ crystal-structure (see **Fig. S1**), phonons, density of states, and charge density are shown in **Fig. 7**. Lower formation energy $Pm$ $(1){-}C_4N_3$ has lower formation energy (-1.03 eV). The main features of $Pm$ $(1){-}C_4N_3$ structure in **Fig. 7a** is the presence of single $C-C$, $C-N$and $N-N$bonds, with fourfold (tetrahedral) coordination of C, and threefold coordination of N. The DFT based Green's function (G) and screened Coulomb interaction (W) approximation [**39**] predicts bandgap of 1.45 eV for $Pm$ $(1){-}C_4N_3$, which lies in the optimal (visible-light absorption) range.

The phonon dispersion and DOS is well proven criteria to establish the structural stability of new materials. The phonon dispersion for $Pm$ $(1){-}C_4N_3$ in **Fig. 7b** is plotted along Γ-Y-C-Z-Γ-B-A-E-D high-symmetry direction in $Pm$ $(1)$ (space group #6) Brillouin zone. The DFT calculated dispersion shows no imaginary modes, i.e., structure is dynamically stable. $Pm$ $(1){-}C_4N_3$ has 56 vibrational modes, we only show some low energy modes, for full dispersion see **Fig. S2**. The phonons within and on boundary correspond to standing waves, where the vibrational modes mainly arise from high symmetry points of the Brillouin zone, e.g., those with $k_1 = k_2 = k_3 = 0$ at Γ-point.

The DFT+HSE calculated partial density of states (PDOS), and $(h,k,l)$-projected charge density for $Pm$ $(1){-}C_4N_3$ are shown in **Fig. 7c** & **7d**. The majority of N-*2p* and C-*2p* states in Fig. 8c are localized near Fermi-level. The nitrogen-*p* and carbon-*p* are the major bonding source in $Pm$ $(1){-}C_4N_3$**,** which originates from the bonding of C-C (*sp3*) and N-C (*sp2*) network and leads to significant hybridization between the two bonding states. The increased stability of *sp3*-*sp2* bonded $Pm\ (1){-}C_4N_3$ comes from the fact from the larger s-contribution as the C-*2s*/N-*2s* are lower in energy than C-*2p*/N-*2p*. We believe that the increased stability arises from the hyperconjugation effect, which is stabilizing interaction resulting mainly from the interaction of *sp3*-bonded networks with *sp2*-bonded networks, which increases the energy stability. In the conduction band (above Fermi-level), both C-*2p* and N-*2p* states show uniform overlap throughout the energy range. The 2D projected charge-density in **Fig. 7d** also reflects the strong directional binding between C and N, which suggest towards the covalent nature of the bonding between C and N.

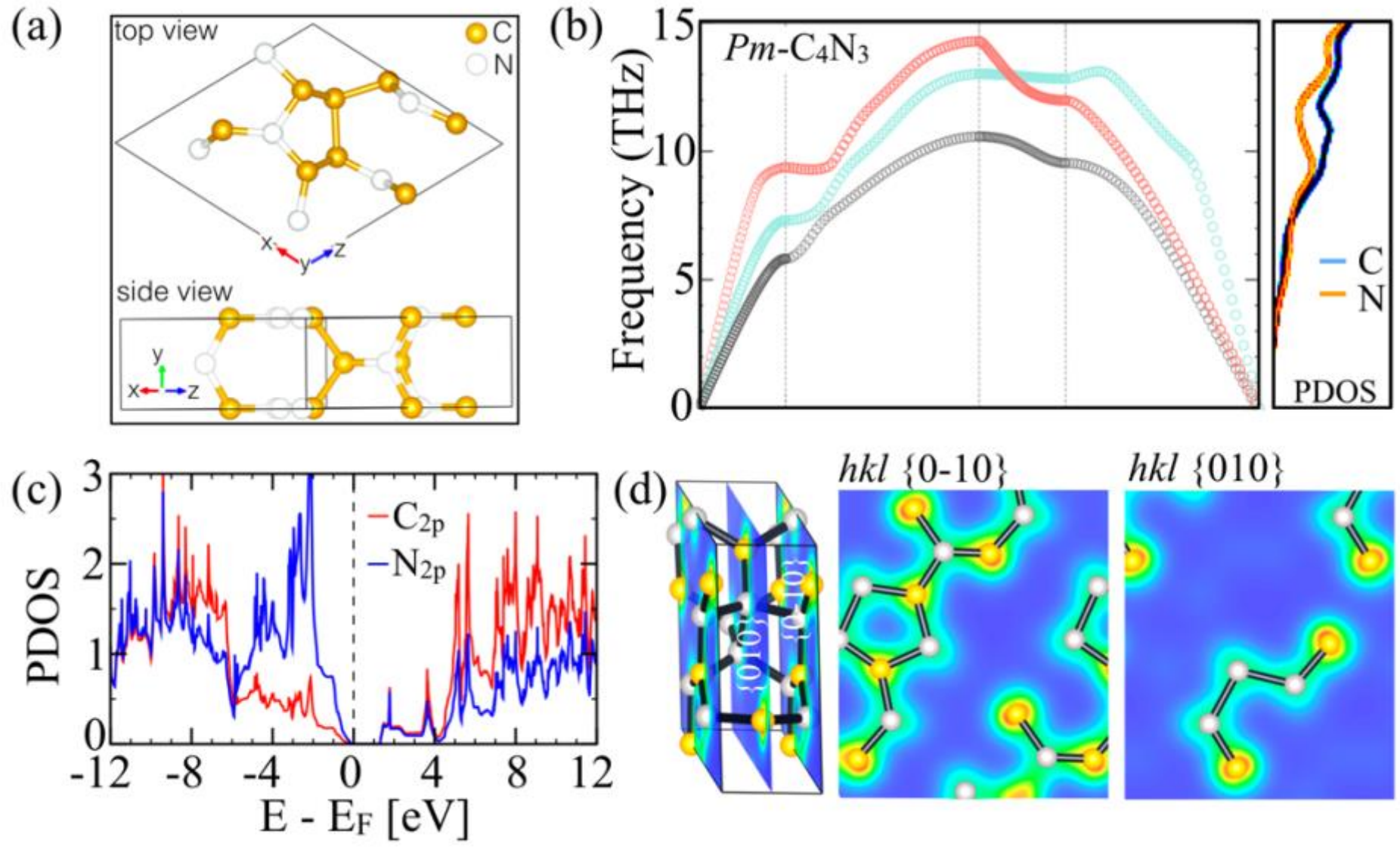

**Figure 7**. (a) Predicted $Pm\ (1){-}C_4N_3$ monoclinic crystal structure with lattice constant a=6.275 Å, b=2.509 Å, c=6.448 Å, and cell angles $\alpha$=90º, $\beta$=116.68º, $\gamma$ =90º. (b) $Pm\ (1){-}C_4N_3$ shows stable phonons at 0 pressure. (see Fig. S2 for full phonon dispersion). (c, d) DFT+HSE calculated partial density of states, and $(h,k,l)$ −projected charge density.

*Pressure dependence of bandgap:* In **Fig. 8**, we show that the bandgap for $Pm\ (1){-}C_4N_3$ does not change much under applied hydrostatic pressure. The bandgap is quite robust as it does not change much under the effect of hydrostatic pressure. The reason for the least effect of pressure on band gap of $Pm\ (1){-}C_4N_3$ comes from the incompressibility of covalently bonded C-C, C-N, and N-N sub-structures. Covalently, bonded C/N are already saturated in terms of charges, therefore does not change much externally applied pressure. Notably, the high bulk and shear moduli in Table 5 indicates the incompressibility of the three (out of four) newly predicted phases.

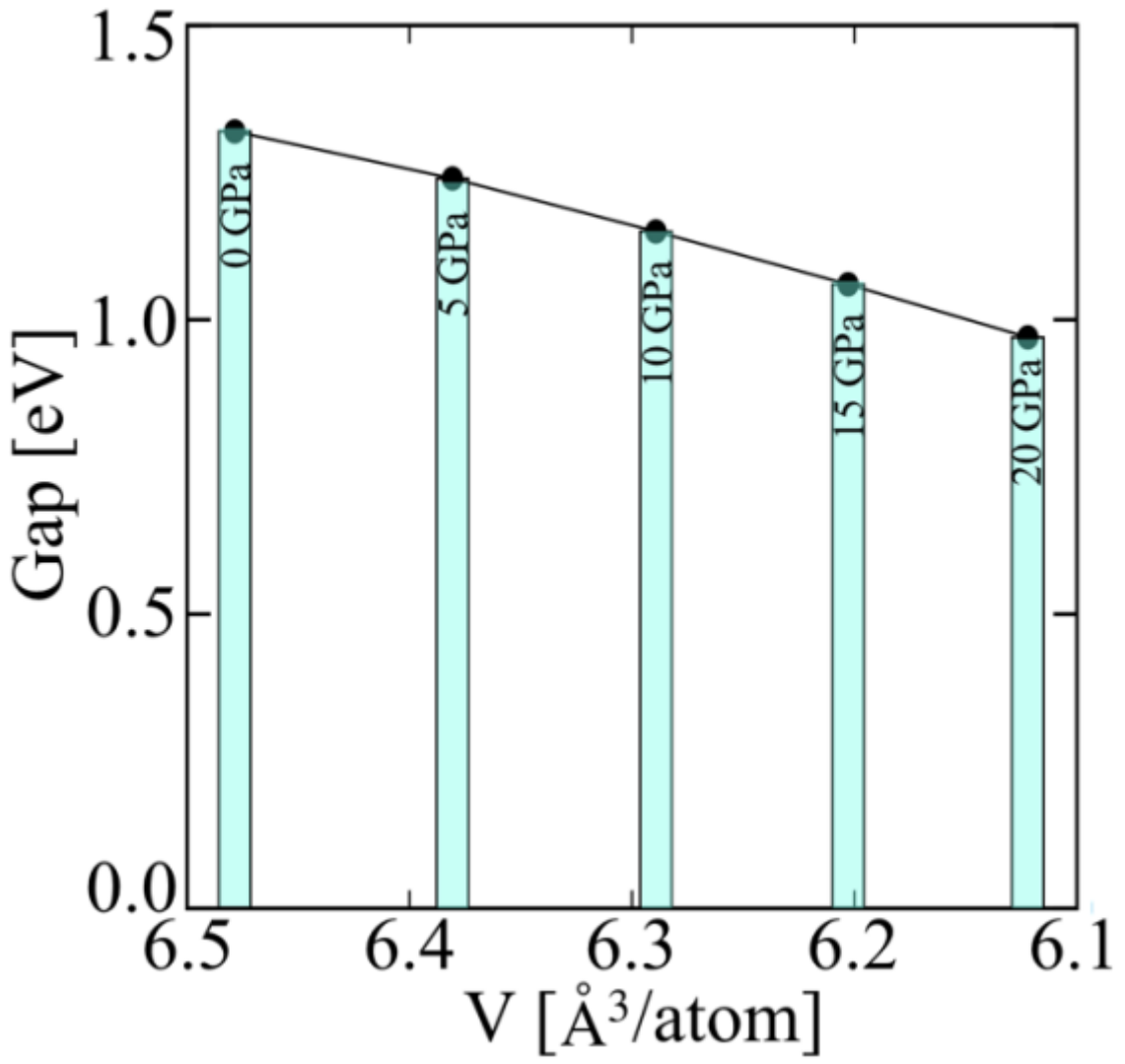

**Figure 8**. The weak change in bandgap under hydrostatic pressure shows highly incompressible nature of $Pm\ (1){-}C_4N_3$.

*Optical property:* Optical absorption in semiconducting materials is very useful especially for photovoltaic and optoelectronic applications, which can be obtained via calculations of the electronic part of the complex dielectric functions. The *GW* method was employed for optical property calculations as it provides more accurate estimate for band gap, band positions, and localized states [**24**]. Two different optical absorption calculation approaches were used for $Pm\ (1) - C_4N_3$: (i) **RPA**: ignores electron-hole interaction, (ii) **BSE**: includes electron-hole interaction. The self-consistent GW [**56**] was done by considering all the valence bands and sufficiently large number of conduction bands.

The GW-RPA and GW-BSE calculated absorption spectra of $Pm\ (1){-}C_4N_3$ are compared, and excitonic peaks are marked by E1, E2 and E3 in **Fig. 9**. The calculated absorption spectra both in GW-RPA ($E_1$=2.6 eV; $E_2$=6.4 eV, $E_3$=12.7 eV) and GW-BSE ($E_1$=4.1 eV; $E_2$=8.3 eV, $E_3$=12.7 eV) show three peak structure. We did not observe any major change in structure of spectra calculated from two different approaches, however, first excitonic peak in GW-RPA is red shifted by ~1.5 eV with respect to GW-BSE. The improvement in GW-BSE calculated absorption spectra comes from the excitonic effects that provides better quasi-particle energies, which is known to strongly affect the optical properties [**24**].

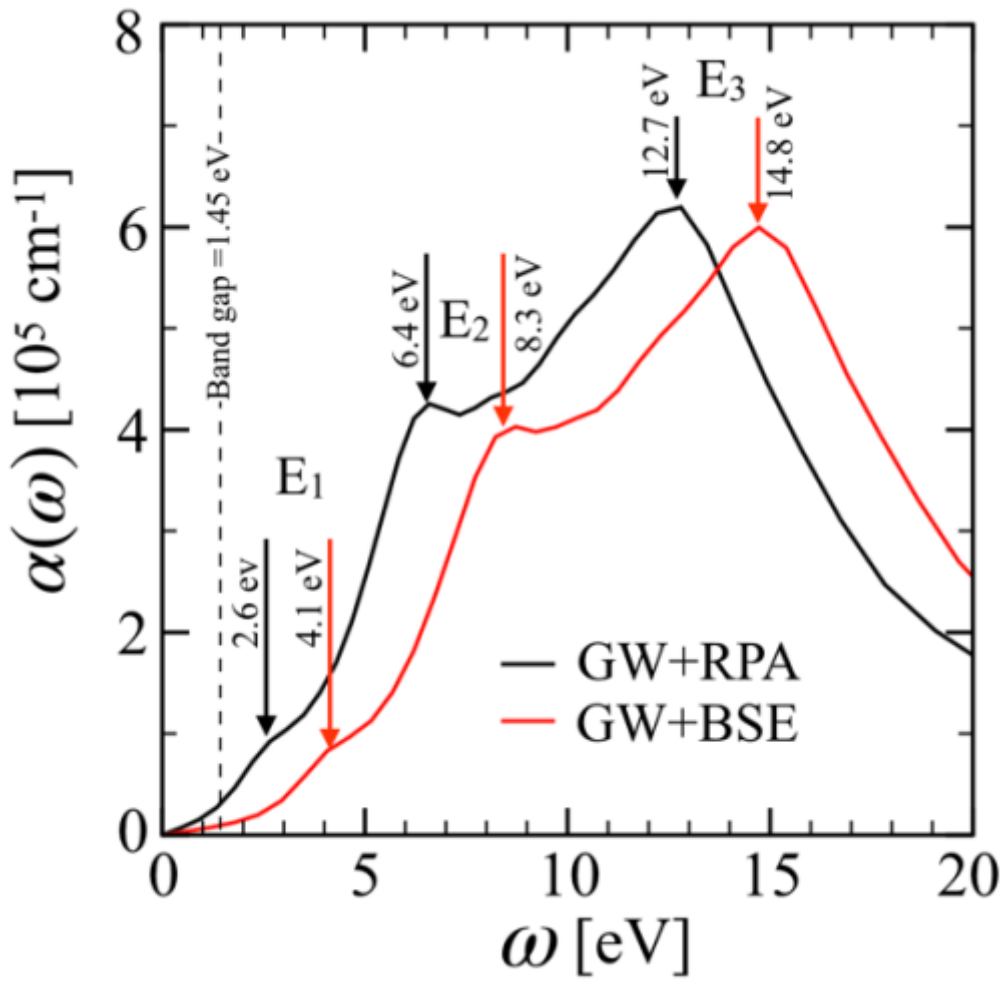

**Figure 9.** Optical absorption in $Pm\ (1){-}C_4N_3$. Imaginary part of macroscopic dielectric function calculated with GW-RPA and GW-BSE. We found that first excitonic peak in GW-RPA is red shifted with respect to GW-BSE by 1.5 eV.

**Discussion and limitation of the proposed metric**: Our results indicate that dynamic stability of covalent systems depends strongly on coordination chemistry and can be described by simple geometric criteria provided by careful consideration given to the limitations. This suggests that similar criteria could be extended to coordination metric definition across the systems. It is reasonable to expect a calibration of the metric value for systems with much different coordination chemistry. Inaccuracy of prediction in systems with similar tetrahedral and trigonal bonding can also be expected to a lesser degree due to the inherently different atomic bond strength and bond lengths that are not accounted for in the simple metric. We feel this does not detract from the utility of this approach as most of the cost in high throughput search methods is spent in ab initio calculation, but a simple rescaling of the metric is cheap, requiring only a few stable structures in a new system to fully describe the characteristics stable coordination weights. Therefore, the user should be careful when using for systems without 3D bonding network, the metric cannot uniquely predict the structural stability. Our future work will focus on generalizing the proposed metric for other bulk inorganic systems beyond 3D bonding network.

## Conclusion

We present a facile `metric' for the identification of structurally (phonons) stable inorganic compounds. Metric uses charge-imbalance within the local substructures of crystalline compounds. The metric was tested for three well studied nitride polymorphs: carbon-nitride, boron-nitride, and silicon-nitride. For carbon-nitride polymorphs, the metric predicts three new structurally stable phases, which is further confirmed by direct phonon calculations. Structurally stable phases also satisfy the thermodynamic (phase stability) and mechanical (Born criteria) stability. The hybrid functional predicted optical band gap and optical property analysis of $Pm\ (1)-C_4N_3$ indicates its usefulness for photovoltaic and optoelectronic applications. This makes the newly predicted carbon-nitride phase very exciting finding.

We believe that the proposed approach can be used to accelerate the search of unknown and unexplored phases of inorganic compounds. This can be important particularly when one wishes to move beyond expensive convex hull determination and instead interested in determining whether a given structure is dynamically stable. The 'metric' can further be modified to consider other chemical systems in which coordination and charge balance are expected to play a role in determining the dynamic stability of the crystalline arrangements.

**Acknowledgement**

We acknowledge support from National Science Foundation through grants no. (DMREF) CMMI-1729350. First-principles calculations were carried out at the Supercomputing Facility at Texas A&M University.

# Metric-driven search for structurally stable inorganic compounds

R. Villarreal[a,b,*], P. Singh[a,*,+], and, R. Arroyave[a,b,c]

[a]*Department of Materials Science & Engineering, Texas A&M University, College Station, TX 77843, USA*
[b] *Department of Mechanical Engineering, Texas A&M University, College Station, TX 77843, USA*
[c] *Department of Industrial and Systems Engineering, Texas A&M University, College Station, TX 77843, USA*

*These authors equally contributed to the execution of the present work
+ Current address**:** Ames Laboratory, U.S. Department of Energy, Iowa State University, Ames, Iowa 50011 USA

**Section S1: Metric for structural stability for all phases in Table 1 and Table 2**:

(a) Charges defined on local substructures for carbon-nitride phases:

#176 stable at 275 Gpa (metric=0.878)

| Index | Element | Ideal coordination | Weight | Charge | Wyckoff Pts | Rank |
|---|---|---|---|---|---|---|
| 0 | C | tetrahedral | 0.898 | 1.48 | x, y, ¼ | 2 |
| 1 | N | trigonal | 0.803 | -1.1 | x, y, ¼ | 2 |
| 3 | N | trigonal | 0.663 | -0.96 | 1/3, 2/3, ¼ | 0 |

#215_stable (metric = 2.467)

| Index | Element | Ideal coordination | Weight | Charge | Wyckoff Pts | Rank |
|---|---|---|---|---|---|---|
| 0 | C | tetrahedral | 0.949 | 1.19 | ½, 0, 0 | 0 |
| 1 | N | trigonal | 0.289 | -0.91 | x, x, x | 1 |

#119 stable (metric= 0.751)

| Index | Element | Ideal coordination | Weight | Charge | Wyckoff Pts | Rank |
|---|---|---|---|---|---|---|
| 0 | C | tetrahedral | 0.934 | 1.21 | 0, ½, ¾ | 0 |
| 1 | N | trigonal | 0.622 | -0.57 | 0, 0, z | 1 |

#122 stable (metric=0.611)

| Index | Element | Ideal coordination | Weight | Charge | Wyckoff Pts | Rank |
|---|---|---|---|---|---|---|
| 0 | C | tetrahedral | 0.872 | 1.26 | x, ¼, 1/8 | 1 |
| 1 | N | trigonal | 0.713 | -0.63 | x, y, z | 3 |

#227_unstable (metric=divergent)

| Index | Element | Ideal coordination | Weight | Charge | Wyckoff Pts | Rank |
|---|---|---|---|---|---|---|
| 0 | C | tetrahedral | 0.958 | 0.95 | 3/8, 3/8, 3/8 | 0 |
| 1 | C | tetrahedral | 0.0 | 0.99 | 0, 0, 0 | 0 |
| 2 | N | trigonal | 0.0 | -0.74 | x, x, x | 1 |

#220 stable (metric=1.109)

| Index | Element | Ideal coordination | Weight | Charge | Wyckoff Pts | Rank |
|---|---|---|---|---|---|---|
| 0 | C | tetrahedral | 0.942 | 1.29 | 3/8, 0, 1/4 | 0 |
| 1 | N | trigonal | 0.637 | -0.98 | x, x, x | 1 |

#159_stable (metric=0.957)

| Index | Element | Ideal coordination | Weight | Charge | Wyckoff Pts | Rank |
|---|---|---|---|---|---|---|
| 0 | C | tetrahedral | 0.829 | 2.01 | x, y, z | 3 |
| 1 | C | tetrahedral | 0.934 | 1.42 | x, y, z | 3 |
| 2 | N | trigonal | 0.657 | -1.05 | 0, 0, z | 1 |
| 3 | N | trigonal | 0.556 | -0.97 | 1/3, 2/3, z | 1 |
| 4 | N | trigonal | 0.75 | -1.06 | x, y, z | 3 |
| 5 | N | trigonal | 0.689 | -1.08 | x, y, z | 3 |

#36_stable (metric=0.623)

| Index | Element | Ideal coordination | Weight | Charge | Wyckoff Pts | Rank |
|---|---|---|---|---|---|---|
| 0 | C | tetrahedral | 0.788 | 1.26 | 0, y, z | 2 |
| 1 | N | trigonal | 0.476 | -0.63 | 0, y, z | 2 |
| 2 | N | trigonal | 0.789 | -0.63 | 0, y, z | 2 |

#225_unstable (metric=divergent)

| Index | Element | Ideal coordination | Weight | Charge | Wyckoff Pts | Rank |
|---|---|---|---|---|---|---|
| 0 | C | tetrahedral | 0 | -0.08 | 1/2, 1/2, 1/2 | 0 |
| 1 | C | tetrahedral | 0 | 0.21 | 1/4, 1/4, 1/4 | 0 |
| 2 | N | trigonal | 0 | -0.34 | 0, 0, 0 | 0 |

#221_unstable (metric=0.594)

| Index | Element | Ideal coordination | Weight | Charge | Wyckoffs | Rank |
|---|---|---|---|---|---|---|
| 0 | C | tetrahedral | 0.725 | 1.92 | 1/2, y, y | 1 |
| 1 | N | trigonal | 0.813 | -2.89 | x, x, x | 1 |

#215 stable (metric=2.467)

| Index | Element | Ideal coordination | Weight | Charge | Wyckoff Pts | Rank |
|---|---|---|---|---|---|---|
| 0 | C | tetrahedral | 0.72 | 1.96 | x, x, z | 2 |
| 1 | N | trigonal | 0.757 | -3.0 | x, x, x | 1 |
| 2 | N | trigonal | 0.857 | -2.87 | x, x, x | 1 |

#227 unstable (Metric = 0.5364)

| Index | Element | Ideal coordination | Weight | Wyckoff Pts | Rank |
|---|---|---|---|---|---|
| 0 | C | tetrahedral | 0.95799 | 3/8, 3/8, 3/8 | 0 |
| 1 | C | tetrahedral | 0.95799 | 3/8, 3/8, 3/8 | 0 |
| 2 | C | tetrahedral | 0.00000 | 0, 0, 0 | 0 |
| 3 | C | tetrahedral | 0.00000 | 0, 0, 0 | 0 |
| 4 | C | tetrahedral | 0.00000 | 0, 0, 0 | 0 |
| 5 | C | tetrahedral | 0.00000 | 0, 0, 0 | 0 |
| 6 | N | trigonal | 0.44635 | x, x, x | 1 |
| 7 | N | trigonal | 0.44635 | x, x, x | 1 |
| 8 | N | trigonal | 0.44635 | x, x, x | 1 |
| 9 | N | trigonal | 0.44635 | x, x, x | 1 |
| 10 | N | trigonal | 0.44635 | x, x, x | 1 |
| 11 | N | trigonal | 0.44635 | x, x, x | 1 |
| 12 | N | trigonal | 0.44635 | x, x, x | 1 |
| 13 | N | trigonal | 0.44635 | x, x, x | 1 |

**(b) Charges defined on local substructures for newly predicted carbon-nitride phases:**

#147 stable (metric=0.843)

| Index | Element | Ideal coordination | Weight | Charge | Wyckoff Pts | Rank |
|---|---|---|---|---|---|---|
| 0 | C | tetrahedral | 0.746 | 0.03 | 1/3, 2/3, z | 1 |
| 1 | C | tetrahedral | 0.758 | -0.08 | 1/3, 2/3, z | 1 |
| 2 | C | tetrahedral | 0.822 | 2.93 | x, y, z | 3 |
| 3 | C | tetrahedral | 0.828 | 2.44 | x, y, z | 3 |
| 4 | N | trigonal | 0.723 | -2.86 | x, y, z | 3 |
| 5 | N | trigonal | 0.713 | -2.48 | x, y, z | 3 |

#6 (2) stable (metric=0.968)

| Index | Element | Ideal coordination | Weight | Charge | Wyckoff Pts | Rank |
|---|---|---|---|---|---|---|
| 0 | C | tetrahedral | 0.775 | -0.08 | x, y, z | 3 |
| 1 | C | tetrahedral | 0.016 | 0.01 | x, y, z | 3 |
| 2 | C | tetrahedral | 0.854 | 2.38 | x, y, z | 3 |
| 3 | C | tetrahedral | 0.862 | 1.28 | x, y, z | 3 |
| 4 | C | tetrahedral | 0.865 | 1.82 | x, y, z | 3 |
| 5 | C | tetrahedral | 0.827 | 1.0 | x, 0, z | 2 |
| 6 | C | tetrahedral | 0.872 | 0.49 | x, 1/2, z | 2 |
| 7 | C | tetrahedral | 0.913 | 2.1 | x, 0, z | 2 |
| 8 | C | tetrahedral | 0.864 | 1.34 | x, 1/2, z | 2 |
| 9 | C | tetrahedral | 0.905 | 2.07 | x, 0, z | 2 |
| 10 | C | tetrahedral | 0.884 | 1.36 | x, 1/2, z | 2 |
| 11 | N | trigonal | 0.733 | -1.13 | x, y, z | 3 |
| 12 | N | trigonal | 0.754 | -1.17 | x, y, z | 3 |
| 13 | N | trigonal | 0.72 | -1.39 | x, 0, z | 2 |
| 14 | N | trigonal | 0.667 | -2.48 | x, 1/2, z | 2 |
| 15 | N | trigonal | 0.743 | -2.83 | x, 0, z | 2 |
| 16 | N | trigonal | 0.761 | -2.38 | x, 1/2, z | 2 |
| 17 | N | trigonal | 0.766 | -2.74 | x, 0, z | 2 |
| 18 | N | trigonal | 0.723 | -1.11 | x, 1/2, z | 2 |
| 19 | N | trigonal | 0.623 | -0.92 | x, 0, z | 2 |
| 20 | N | trigonal | 0.613 | -0.84 | x, 1/2, z | 2 |

#11 stable (metric=0.289)

| Index | Element | Ideal coordination | Weight | Charge | Wyckoff Pts | Rank |
|---|---|---|---|---|---|---|
| 0 | C | tetrahedral | 0.0 | 0.29 | x, 1/4, z | 2 |
| 1 | C | tetrahedral | 0.806 | 0.84 | x, 1/4, z | 2 |
| 2 | C | tetrahedral | 0.007 | 3.29 | x, 1/4, z | 2 |
| 3 | C | tetrahedral | 0.007 | 1.64 | x, 1/4, z | 2 |
| 4 | N | trigonal | 0.344 | -0.98 | x, 1/4, z | 2 |
| 5 | N | trigonal | 0.85 | -2.68 | x, 1/4, z | 2 |
| 6 | N | trigonal | 0.402 | -2.4 | x, 1/4, z | 2 |

**(c) Charges defined on local substructures of boron-nitride phases:**

#194, unstable (metric = divergent)

| Index | Element | Ideal coordination | Weight | Charge | Wyckoff Pts | Rank |
|---|---|---|---|---|---|---|
| 0 | B | tetrahedral | 0.0 | -3.0 | 1/3, 2/3, 1/4 | 0 |
| 1 | B | tetrahedral | 0.0 | -3.0 | 1/3, 2/3, 1/4 | 0 |
| 2 | N | trigonal | 0.892 | 3.0 | 1/3, 2/3, 3/4 | 0 |
| 3 | N | trigonal | 0.892 | 3.0 | 1/3, 2/3, 3/4 | 0 |

#14, unstable (metric = 0.0254)

| Index | Element | Ideal coordination | Weight | Charge | Wyckoff Pts | Rank |
|---|---|---|---|---|---|---|
| 0 | B | tetrahedral | 0.005 | -3.0 | x, y, z | 3 |
| 1 | B | tetrahedral | 0.005 | -3.0 | x, y, z | 3 |
| 2 | B | tetrahedral | 0.005 | -3.0 | x, y, z | 3 |
| 3 | B | tetrahedral | 0.005 | -3.0 | x, y, z | 3 |
| 4 | N | trigonal | 0.19 | 3.0 | x, y, z | 3 |
| 5 | N | trigonal | 0.19 | 3.0 | x, y, z | 3 |
| 6 | N | trigonal | 0.19 | 3.0 | x, y, z | 3 |
| 7 | N | trigonal | 0.19 | 3.0 | x, y, z | 3 |

#187, unstable (metric = 0.0)

| Index | Element | Ideal coordination | Weight | Charge | Wyckoff Pts | Rank |
|---|---|---|---|---|---|---|
| 0 | B | tetrahedral | 0.0 | -3.0 | 0, 0, 0 | 0 |
| 1 | B | tetrahedral | 0.0 | -3.0 | 1/3, 2/3, 1/2 | 0 |
| 2 | N | trigonal | 0.91 | 2.999 | 1/3, 2/3, 0 | 0 |
| 3 | N | trigonal | 0.972 | 3.001 | 2/3, 1/3, 1/2 | 0 |

#62, stable 0(metric = 7.2175)

| Index | Element | Ideal coordination | Weight | Charge | Wyckoff Pts | Rank |
|---|---|---|---|---|---|---|
| 0 | B | tetrahedral | 0.57 | -3.0 | x, 1/4, z | 2 |
| 1 | B | tetrahedral | 0.57 | -3.0 | x, 1/4, z | 2 |
| 2 | B | tetrahedral | 0.57 | -3.0 | x, 1/4, z | 2 |
| 3 | B | tetrahedral | 0.57 | -3.0 | x, 1/4, z | 2 |
| 4 | N | trigonal | 0.079 | 3.0 | x, 1/4, z | 2 |
| 5 | N | trigonal | 0.079 | 3.0 | x, 1/4, z | 2 |
| 6 | N | trigonal | 0.079 | 3.0 | x, 1/4, z | 2 |
| 7 | N | trigonal | 0.079 | 3.0 | x, 1/4, z | 2 |

**(d) Charges defined on local substructures of silicon-nitride polymorphs:**

#159, unstable (metric = 0.4918)

| Index | Element | Ideal coordination | Weight | Charge | Wyckoff Pts | Rank |
|---|---|---|---|---|---|---|
| 0 | Si | tetrahedral | 0.25691 | -3.39059 | x, y, z | 3 |
| 1 | Si | tetrahedral | 0.25691 | -3.38440 | x, y, z | 3 |
| 2 | Si | tetrahedral | 0.25691 | -3.40369 | x, y, z | 3 |
| 3 | Si | tetrahedral | 0.01288 | -4.00000 | x, y, z | 3 |
| 4 | Si | tetrahedral | 0.01288 | -4.00000 | x, y, z | 3 |
| 5 | Si | tetrahedral | 0.01288 | -4.00000 | x, y, z | 3 |
| 6 | Si | tetrahedral | 0.25691 | -3.40369 | x, y, z | 3 |
| 7 | Si | tetrahedral | 0.25691 | -3.38440 | x, y, z | 3 |
| 8 | Si | tetrahedral | 0.25691 | -3.39059 | x, y, z | 3 |
| 9 | Si | tetrahedral | 0.01288 | -4.00000 | x, y, z | 3 |
| 10 | Si | tetrahedral | 0.01288 | -4.00000 | x, y, z | 3 |
| 11 | Si | tetrahedral | 0.01288 | -4.00000 | x, y, z | 3 |
| 12 | N | trigonal | 0.41533 | 2.97835 | x, y, z | 3 |
| 13 | N | trigonal | 0.13318 | 3.05646 | x, y, z | 3 |
| 14 | N | trigonal | 0.41533 | 2.97835 | x, y, z | 3 |
| 15 | N | trigonal | 0.13318 | 3.05777 | x, y, z | 3 |
| 16 | N | trigonal | 0.41534 | 2.97509 | x, y, z | 3 |
| 17 | N | trigonal | 0.13318 | 3.05777 | x, y, z | 3 |
| 18 | N | trigonal | 0.00000 | 2.08076 | 0, 0, z | 1 |
| 19 | N | trigonal | 0.41534 | 2.97624 | x, y, z | 3 |
| 20 | N | trigonal | 0.13318 | 3.05646 | x, y, z | 3 |
| 21 | N | trigonal | 0.00000 | 2.00115 | 0, 0, z | 1 |
| 22 | N | trigonal | 0.41534 | 2.97624 | x, y, z | 3 |
| 23 | N | trigonal | 0.13318 | 3.05288 | x, y, z | 3 |
| 24 | N | trigonal | 0.41534 | 2.97509 | x, y, z | 3 |
| 25 | N | trigonal | 0.13318 | 3.05288 | x, y, z | 3 |
| 26 | N | trigonal | 0.00000 | 2.08076 | 0, 0, z | 1 |
| 27 | N | trigonal | 0.00000 | 2.00114 | 0, 0, z | 1 |

#176, stable (metric = 1.1767)

| Index | Element | Ideal coordination | Weight | Charge | Wyckoff Pts | Rank |
|---|---|---|---|---|---|---|
| 0 | Si | tetrahedral | 0.69308 | -4.00000 | x, y, z | 3 |
| 1 | Si | tetrahedral | 0.81228 | -4.00000 | x, y, 1/4 | 2 |
| 2 | Si | tetrahedral | 0.81228 | -4.00000 | x, y, 1/4 | 2 |
| 3 | Si | tetrahedral | 0.81228 | -4.00000 | x, y, 1/4 | 2 |
| 4 | Si | tetrahedral | 0.69308 | -4.00000 | x, y, z | 3 |
| 5 | Si | tetrahedral | 0.81228 | -4.00000 | x, y, 1/4 | 2 |
| 6 | Si | tetrahedral | 0.69308 | -4.00000 | x, y, z | 3 |
| 7 | Si | tetrahedral | 0.81228 | -4.00000 | x, y, 1/4 | 2 |
| 8 | Si | tetrahedral | 0.69308 | -4.00000 | x, y, z | 3 |
| 9 | Si | tetrahedral | 0.69308 | -4.00000 | x, y, z | 3 |
| 10 | Si | tetrahedral | 0.69308 | -4.00000 | x, y, z | 3 |
| 11 | Si | tetrahedral | 0.69308 | -4.00000 | x, y, z | 3 |
| 12 | Si | tetrahedral | 0.69308 | -4.00000 | x, y, z | 3 |
| 13 | Si | tetrahedral | 0.81228 | -4.00000 | x, y, 1/4 | 2 |
| 14 | Si | tetrahedral | 0.69308 | -4.00000 | x, y, z | 3 |
| 15 | Si | tetrahedral | 0.69308 | -4.00000 | x, y, z | 3 |
| 16 | Si | tetrahedral | 0.69308 | -4.00000 | x, y, z | 3 |
| 17 | Si | tetrahedral | 0.69308 | -4.00000 | x, y, z | 3 |
| 18 | N | trigonal | 0.22775 | 2.97236 | x, y, 1/4 | 2 |
| 19 | N | trigonal | 0.42526 | 2.98976 | x, y, 1/4 | 2 |
| 20 | N | trigonal | 0.60735 | 3.03001 | x, y, z | 3 |
| 21 | N | trigonal | 0.22775 | 2.97236 | x, y, 1/4 | 2 |
| 22 | N | trigonal | 0.22775 | 2.97862 | x, y, 1/4 | 2 |
| 23 | N | trigonal | 0.60827 | 3.01866 | x, y, z | 3 |
| 24 | N | trigonal | 0.60827 | 3.01123 | x, y, z | 3 |
| 25 | N | trigonal | 0.42526 | 3.00352 | x, y, 1/4 | 2 |
| 26 | N | trigonal | 0.42526 | 2.98975 | x, y, 1/4 | 2 |
| 27 | N | trigonal | 0.42526 | 3.00352 | x, y, 1/4 | 2 |
| 28 | N | trigonal | 0.60735 | 3.03144 | x, y, z | 3 |
| 29 | N | trigonal | 0.22775 | 2.97859 | x, y, 1/4 | 2 |
| 30 | N | trigonal | 0.60735 | 3.02984 | x, y, z | 3 |
| 31 | N | trigonal | 0.22775 | 2.97530 | x, y, 1/4 | 2 |
| 32 | N | trigonal | 0.60735 | 3.03160 | x, y, z | 3 |
| 33 | N | trigonal | 0.60735 | 3.00027 | x, y, z | 3 |
| 34 | N | trigonal | 0.60827 | 3.02968 | x, y, z | 3 |
| 35 | N | trigonal | 0.60827 | 3.00013 | x, y, z | 3 |
| 36 | N | trigonal | 0.60735 | 3.00196 | x, y, z | 3 |
| 37 | N | trigonal | 0.60735 | 3.00028 | x, y, z | 3 |
| 38 | N | trigonal | 0.60735 | 3.00176 | x, y, z | 3 |
| 39 | N | trigonal | 0.22775 | 2.97546 | x, y, 1/4 | 2 |
| 40 | N | trigonal | 0.42526 | 2.98695 | x, y, 1/4 | 2 |
| 41 | N | trigonal | 0.42526 | 2.98697 | x, y, 1/4 | 2 |

#62, stable (metric = 3.4691)

| Index | Element | Ideal coordination | Weight | Charge | Wyckoff Pts | Rank |
|---|---|---|---|---|---|---|
| 0 | Si | tetrahedral | 0.581 | -4.0 | x, 1/4, z | 2 |
| 1 | Si | tetrahedral | 0.581 | -4.0 | x, 1/4, z | 2 |
| 2 | Si | tetrahedral | 0.049 | -4.0 | x, 1/4, z | 2 |
| 3 | Si | tetrahedral | 0.581 | -4.0 | x, 1/4, z | 2 |
| 4 | Si | tetrahedral | 0.049 | -4.0 | x, 1/4, z | 2 |
| 5 | Si | tetrahedral | 0.581 | -4.0 | x, 1/4, z | 2 |
| 6 | Si | tetrahedral | 0.049 | -4.0 | x, 1/4, z | 2 |
| 7 | Si | tetrahedral | 0.0 | -4.0 | 0, 0, 1/2 | 0 |
| 8 | Si | tetrahedral | 0.0 | -4.0 | 0, 0, 1/2 | 0 |
| 9 | Si | tetrahedral | 0.049 | -4.0 | x, 1/4, z | 2 |
| 10 | Si | tetrahedral | 0.0 | -4.0 | 0, 0, 1/2 | 0 |
| 11 | Si | tetrahedral | 0.0 | -4.0 | 0, 0, 1/2 | 0 |
| 12 | N | trigonal | 0.026 | 3.1 | x, 1/4, z | 2 |
| 13 | N | trigonal | 0.026 | 3.101 | x, 1/4, z | 2 |
| 14 | N | trigonal | 0.078 | 2.962 | x, y, z | 3 |
| 15 | N | trigonal | 0.0 | 3.003 | x, 1/4, z | 2 |
| 16 | N | trigonal | 0.078 | 2.962 | x, y, z | 3 |
| 17 | N | trigonal | 0.078 | 2.935 | x, y, z | 3 |
| 18 | N | trigonal | 0.078 | 2.962 | x, y, z | 3 |
| 19 | N | trigonal | 0.0 | 3.003 | x, 1/4, z | 2 |
| 20 | N | trigonal | 0.078 | 2.935 | x, y, z | 3 |
| 21 | N | trigonal | 0.078 | 2.935 | x, y, z | 3 |
| 22 | N | trigonal | 0.0 | 3.003 | x, 1/4, z | 2 |
| 23 | N | trigonal | 0.078 | 2.962 | x, y, z | 3 |
| 24 | N | trigonal | 0.026 | 3.101 | x, 1/4, z | 2 |
| 25 | N | trigonal | 0.0 | 3.003 | x, 1/4, z | 2 |
| 26 | N | trigonal | 0.078 | 2.935 | x, y, z | 3 |
| 27 | N | trigonal | 0.026 | 3.1 | x, 1/4, z | 2 |

#176, stable 9(metric = 0.8281)

| Index | Element | Ideal coordination | Weight | Charge | Wyckoff Pts | Rank |
|---|---|---|---|---|---|---|
| 0 | Si | tetrahedral | 0.866 | -4.0 | x, y, 1/4 | 2 |
| 1 | Si | tetrahedral | 0.866 | -4.0 | x, y, 1/4 | 2 |
| 2 | Si | tetrahedral | 0.866 | -4.0 | x, y, 1/4 | 2 |
| 3 | Si | tetrahedral | 0.866 | -4.0 | x, y, 1/4 | 2 |
| 4 | Si | tetrahedral | 0.866 | -4.0 | x, y, 1/4 | 2 |
| 5 | Si | tetrahedral | 0.866 | -4.0 | x, y, 1/4 | 2 |
| 6 | N | trigonal | 0.681 | 2.999 | 2/3, 1/3, 1/4 | 0 |
| 7 | N | trigonal | 0.681 | 2.999 | 2/3, 1/3, 1/4 | 0 |
| 8 | N | trigonal | 0.818 | 2.998 | x, y, 1/4 | 2 |
| 9 | N | trigonal | 0.818 | 3.004 | x, y, 1/4 | 2 |
| 10 | N | trigonal | 0.818 | 2.999 | x, y, 1/4 | 2 |
| 11 | N | trigonal | 0.818 | 2.999 | x, y, 1/4 | 2 |
| 12 | N | trigonal | 0.818 | 2.998 | x, y, 1/4 | 2 |
| 13 | N | trigonal | 0.818 | 3.004 | x, y, 1/4 | 2 |

**Section S2: Lattice parameters and atomic positions of newly predicted phases of carbon-nitride.**

**Table S1**. Atomic position of predicted structure $Pm\ (1){-}C_4N_3$ with Z=2 of space group Pm (6). Z is the unit cell multiplicity. The lattice constants a=6.27489 Å, b=2.50919 Å, c=6.44779 Å, and angle α=90°, β=116.68°, γ=90°.

| **Atoms** | **Wycoff symmetry** | **Atomic coordinates** | | |
|---|---|---|---|---|
| | | **x** | **y** | **z** |
| C | 1a | 0.77869 | 0.00000 | 0.33078 |
| C | 1a | 0.62533 | 0.50000 | 0.74941 |
| C | 1a | 0.36934 | 0.50000 | 0.51643 |
| C | 1a | 0.74401 | 0.00000 | 0.70115 |
| C | 1a | 0.03637 | 0.00000 | 0.54409 |
| C | 1a | 0.55955 | 0.50000 | 0.95868 |
| C | 1a | 0.38485 | 0.00000 | 0.38102 |
| C | 1a | 0.20589 | 0.00000 | 0.96333 |
| N | 1a | 0.75697 | 0.50000 | 0.19117 |
| N | 1a | 0.41656 | 0..00000 | 0.92760 |
| N | 1a | 0.63617 | 0.00000 | 0.44645 |
| N | 1a | 0.99746 | 0.00000 | 0.75432 |
| N | 1a | 0.15704 | 0.50000 | 0.53744 |
| N | 1a | 0.19726 | 0.00000 | 0.15589 |

**Table S2**. Atomic position of predicted structure $Pm\ (2){-}C_4N_3$ with Z=4 of space group #6. Z is the unit cell multiplicity. The lattice constants a= 6.53931Å, b= 4.80944 Å, c= 6.55202 Å, and angle α=90°, β=119.5°, γ= 90°.

| Atoms | Wycoff symmetry | Atomic coordinates | | |
|---|---|---|---|---|
| | | x | y | z |
| C | c | 0.165501 | 0.085967 | 0.331182 |
| C | c | 0.165505 | 0.41404 | 0.331183 |
| C | c | 0.539665 | 0.994673 | 0.323827 |
| C | c | 0.539685 | 0.505336 | 0.323843 |
| C | c | 0.320973 | 0.00545 | 0.599358 |
| C | c | 0.320973 | 0.49453 | 0.599356 |
| C | c | 0.280076 | 0.00415 | 0.17533 |
| C | c | 0.280065 | 0.495857 | 0.175352 |
| C | c | 0.900918 | 0.00477 | 0.221556 |
| C | c | 0.900926 | 0.495227 | 0.221571 |
| C | a | 0.679076 | 0.250009 | 0.410204 |
| C | b | 0.683975 | 0.750003 | 0.39569 |
| C | a | 0.723617 | 0.249994 | 0.824635 |
| C | b | 0.719044 | 0.749993 | 0.818258 |
| C | a | 0.098718 | 0.250004 | 0.778654 |
| C | b | 0.096706 | 0.750004 | 0.782122 |
| N | c | 0.171293 | 0.997481 | 0.710447 |
| N | c | 0.171282 | 0.502515 | 0.710438 |
| N | c | 0.782569 | 0.997411 | 0.962629 |
| N | c | 0.782576 | 0.502577 | 0.96263 |
| N | a | 0.817284 | 0.250006 | 0.291348 |
| N | b | 0.86267 | 0.750008 | 0.311987 |
| N | a | 0.471284 | 0.24999 | 0.677767 |
| N | b | 0.455376 | 0.749986 | 0.651876 |
| N | a | 0.213702 | 0.250008 | 0.026243 |
| N | b | 0.186 | 0.750003 | 0.040204 |
| N | a | 0.837125 | 0.249993 | 0.674149 |

**Table S3**. Atomic position of predicted structure $P\bar{3}-C_4N_3$ with Z=4 of space group #147. Z is the unit cell multiplicity. The lattice constants a=*6.62865Å*, b= *6.62866 Å*, c= *4.72293Å,* and angle $\alpha$=*90º*, $\beta$=*90º*, $\gamma$= *120º*.

| Atoms | Wycoff symmetry | Atomic coordinates | | |
|---|---|---|---|---|
| | | x | y | z |
| C | d | 0.666666 | 0.333334 | 0.291188 |
| C | d | 0.666666 | 0.333334 | 0.959226 |
| C | d | 0.333335 | 0.666666 | 0.208806 |
| C | d | 0.333335 | 0.666666 | 0.540784 |
| C | g | 0.397178 | 0.220978 | 0.36939 |
| C | g | 0.398723 | 0.219449 | 0.880354 |
| C | g | 0.823801 | 0.602822 | 0.36939 |
| C | g | 0.820727 | 0.601276 | 0.880354 |
| C | g | 0.77902 | 0.176201 | 0.369391 |
| C | g | 0.780549 | 0.179275 | 0.880354 |
| C | g | 0.602753 | 0.778969 | 0.130603 |
| C | g | 0.601207 | 0.780487 | 0.619649 |
| C | g | 0.176216 | 0.397248 | 0.130603 |
| C | g | 0.179281 | 0.398795 | 0.619649 |
| C | g | 0.221034 | 0.823783 | 0.130603 |
| C | g | 0.219516 | 0.820718 | 0.619648 |
| N | g | 0.281348 | 0.968241 | 0.377741 |
| N | g | 0.29925 | 0.95448 | 0.873143 |
| N | g | 0.6869 | 0.718658 | 0.377741 |
| N | g | 0.655236 | 0.700754 | 0.873143 |
| N | g | 0.031758 | 0.313106 | 0.377741 |
| N | g | 0.045519 | 0.344768 | 0.873143 |
| N | g | 0.718967 | 0.031757 | 0.122259 |
| N | g | 0.700993 | 0.045498 | 0.626861 |
| N | g | 0.312783 | 0.281028 | 0.122259 |
| N | g | 0.344502 | 0.299005 | 0.626861 |
| N | g | 0.968246 | 0.687212 | 0.122259 |

**Table S4**. Atomic position of predicted structure $P2_1/m{-}C_4N_3$ with Z=2 of space group #11. Z is the unit cell multiplicity. The lattice constants a= *8.72846 Å*, b= *6.0239 Å*, c= *2.35129 Å*, and angle $\alpha$*=90º*, $\beta$*=90º*, $\gamma$*= 114.7º*.

| Atoms | Wycoff symmetry | Atomic coordinates | | |
|---|---|---|---|---|
| | | x | y | z |
| C | 1e | 0.685822 | 0.75 | 0.133944 |
| C | 1e | 0.566044 | 0.25 | 0.866361 |
| C | 1e | 0.433956 | 0.75 | 0.133639 |
| C | 1e | 0.724889 | 0.75 | 0.66663 |
| C | 1e | 0.275111 | 0.25 | 0.33337 |
| C | 1e | 0.909204 | 0.75 | 0.25824 |
| C | 1e | 0.090796 | 0.25 | 0.74176 |
| C | 1e | 0.314178 | 0.25 | 0.866056 |
| N | 1e | 0.686954 | 0.75 | 0.956625 |
| N | 1e | 0.038391 | 0.25 | 0.287881 |
| N | 1e | 0.961609 | 0.75 | 0.712119 |
| N | 1e | 0.611104 | 0.25 | 0.688953 |
| N | 1e | 0.388896 | 0.75 | 0.311047 |
| N | 1e | 0.313046 | 0.25 | 0.043375 |

**Section S3: Pm-$C_4N_3$ crystal structure and theoretical powder diffraction**:

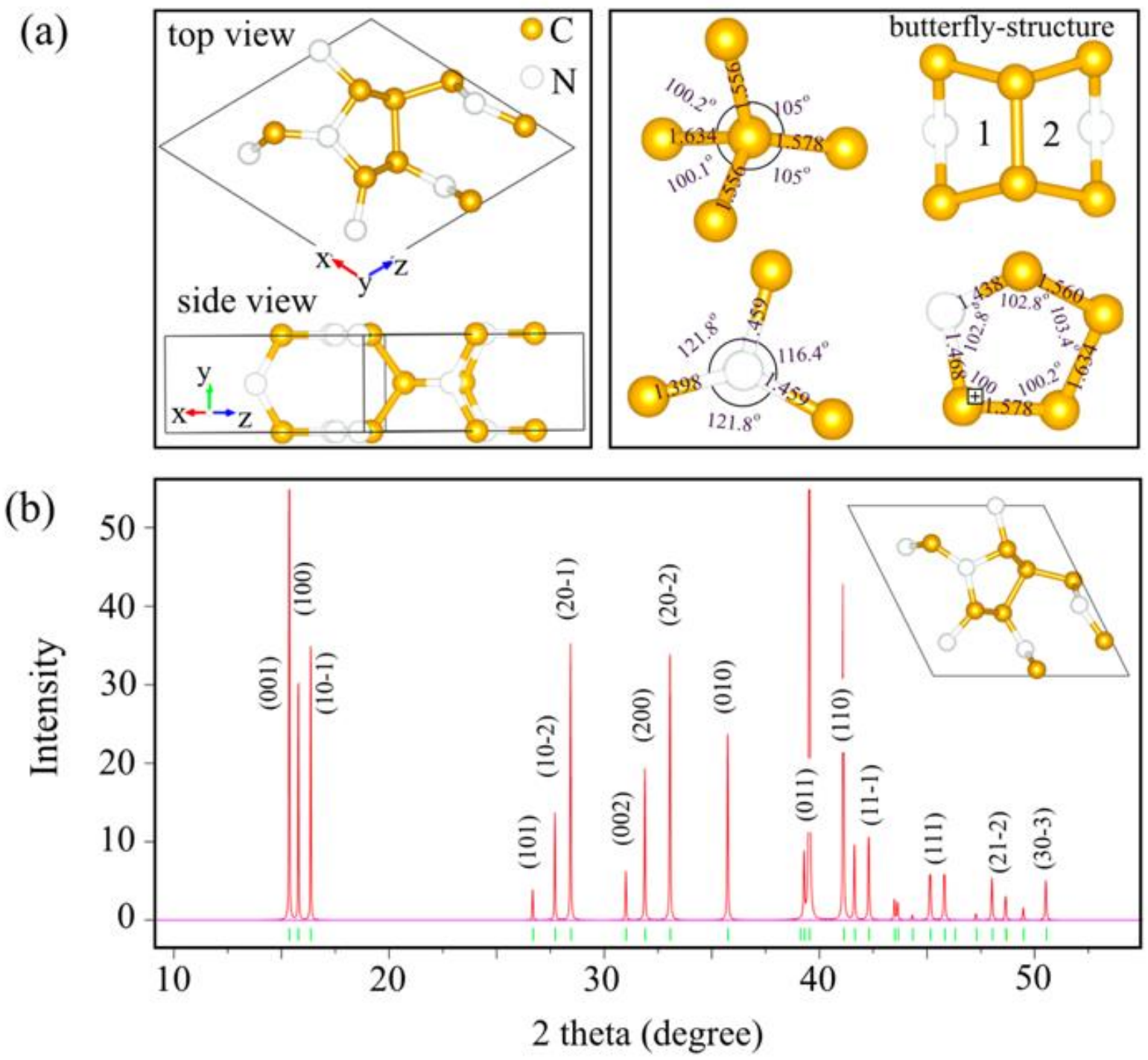

**Figure S1**. (a) Crystal-structure, bond-length, and bond-angles for DFT, and (b) theoretically predicted powder diffraction pattern.

**Full phonon dispersion for $\boldsymbol{Pm\,(1)-C_4N_3}$: Figure. S3** shows the phonon dispersion curves of $Pm\,(1)-C_4N_3$ along Γ-Y-C-Z-Γ-B-A-E-D (space group # 6) high symmetry lines and partial phonon density of states. At high frequency region (about 50 THz), the phonon branch shows weak dispersive relation compared with the optical phonon branches in the low frequency regions. The phonon density of states shows In The low frequency region (<40 THz), the partial phonon density of states is mainly from the vibration of C or N atoms. The phonon properties (force-constants) are the prerequisite for the thermo-dynamical and thermal expansion properties.

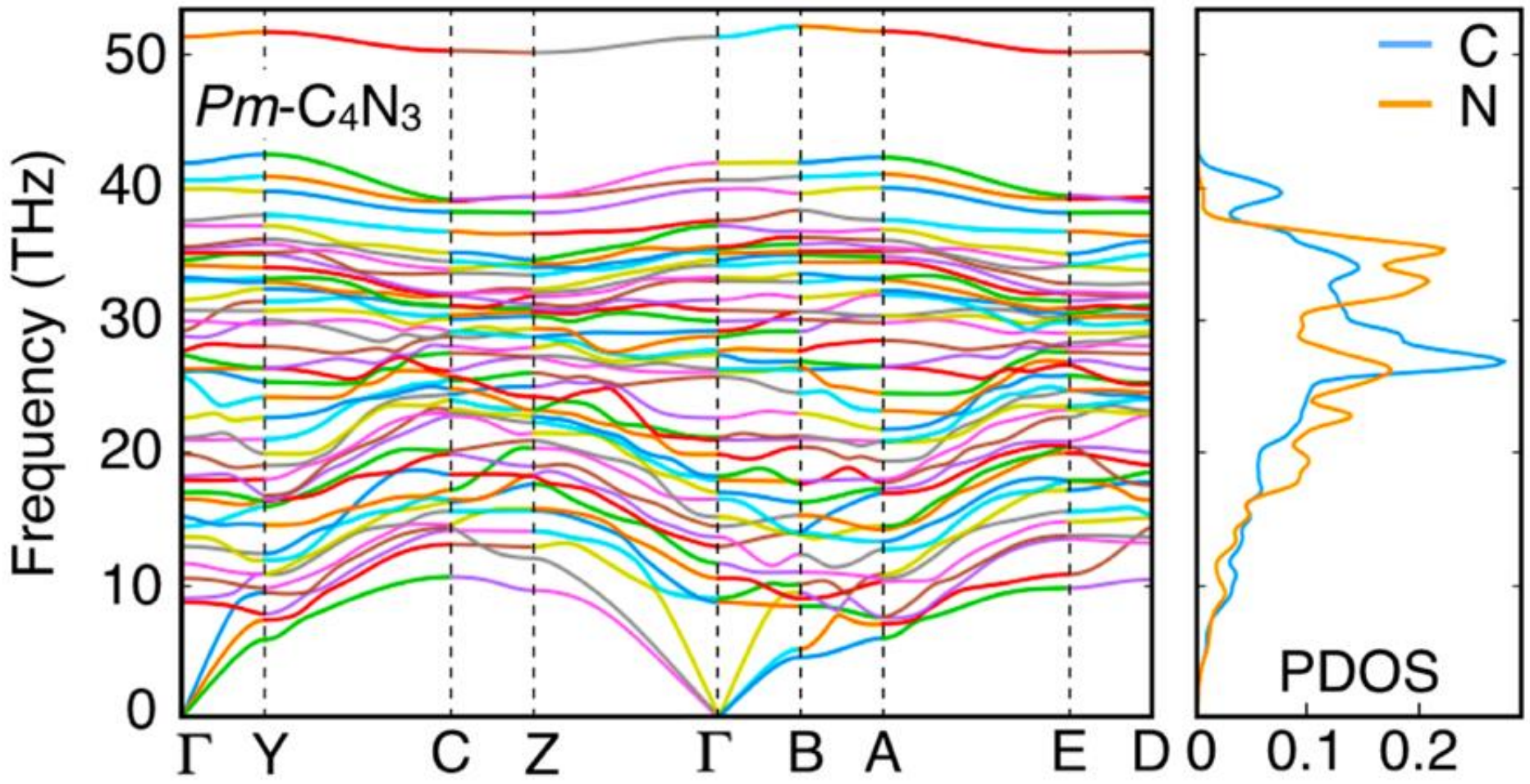

**Figure S2**. Calculated phonon dispersion along high-symmetry direction in Brillouin zone and partial phonon density of states for $Pm\,(1)-C_4N_3$.